\documentclass[aps,prd,onecolumn,preprintnumbers,nofootinbib,amsmath,amssymb]{revtex4-2}

\usepackage{graphicx} 
\usepackage{bm}
\usepackage{hyperref}
\usepackage{cleveref}
\usepackage{float}
\usepackage{subcaption}
\graphicspath{{Figuras/}}
\begin{document}

% Top right preprint numbers
%\preprint{a}

\title{Cosmology in sixth-order higher-derivative theories of gravity}

% Authors and Affiliations
\author{Carlos Aguilar}
\email{c.aguilar@unizar.es} 
\author{Manuel Asorey} % Changed to Autor2 for clarity
\email{asorey@unizar.es} 
\author{Diego Muñoz} % Changed to Autor2 for clarity
\email{diego.munoz@unizar.es} 
\author{Miguel Pardina} % Changed to Autor2 for clarity
\email{mpardina@unizar.es} 
\affiliation{Centro de Astropart\'{\i}culas y F\'{\i}sica de Altas Energ\'{\i}as, Departamento de F\'{\i}sica Te\'orica.\break
Universidad de Zaragoza,  E-50009 Zaragoza, Spain}

\date{\today} % \today is a helpful default command here

\begin{abstract}
We analyze the cosmological implications of super-renormalizable theories of gravity with  quadratic curvature invariants $R^2$ and $C^{\mu\nu\rho\sigma}C_{\mu\nu\rho\sigma}$, and  six-derivative operators $R\Box R$ and $C^{\mu\nu\rho\sigma}\Box C_{\mu\nu\rho\sigma}$. We first analyze homogeneous and isotropic cosmological solutions and show that the theory admits a background dynamics richer than quadratic gravity, including expanding, recollapsing and non-singular bouncing solutions. We clarify in the Einstein-frame formulation the  mechanism that circumvents the Hawking--Penrose singularity theorems, and identify the crucial role played by the additional higher-derivative degrees of freedom in that mechanism. We show that the  complete set of linear equations governing tensor perturbations on Minkowski, radiation-dominated, matter-dominated and de Sitter backgrounds, admits analytically stable solutions. The analysis reveals that the stability properties are determined by the pole structure of the propagator and by specific relations among the higher-derivative couplings, while the singularity-free backgrounds are found to be unstable against tensor perturbations. In the case of scalar cosmological perturbations  the higher-derivative theories present unavoidable super-horizon scalar instabilities in de Sitter spacetimes for both quadratic and sixth-order gravity. Therefore, to extract inflationary observables, we adopt an Effective Field Theory (EFT) approach to the sixth-order theory, allowing the higher-derivative operators to be treated consistently as perturbative corrections around the Starobinsky inflationary background. Within this framework, we compute the modifications of the primordial scalar and tensor power spectra and determine the resulting predictions for the scalar spectral index and tensor-to-scalar ratio. We find that sixth-order operators produce controlled deviations from the Starobinsky model while preserving the effective field theory expansion, thereby providing a consistent phenomenological framework to probe ultraviolet gravitational corrections through cosmological observables.
\end{abstract}

% This command is required to generate the title block!
\maketitle

% You need at least some body text to generate a page.
\section{Introduction}

Einstein's General Theory of Relativity (GR) is still the most successful theory for explaining the gravitational phenomena since it has passed numerous experimental tests to date in its domain, whose range of scales goes between the micrometers up to the large-scale of the universe \cite{Will:2001mx}. However, there are several conceptual and observational challenges with our current understanding of GR and its corresponding Lambda cold dark matter ($\Lambda$CDM) cosmological model  of the Universe evolution \cite{Weinberg:1988cp,Perivolaropoulos:2021jda}.

From a fundamental physics perspective, GR is the classical theory of spacetime whose dynamics is governed by the Einstein-Hilbert (EH) action. The spacetime solutions of the theory contain singularities, for instance inside black holes and at the beginning of the universe \cite{Hawking:1970zqf}, leading to a complete breakdown of predictability in those scenarios. Moreover, in the framework of Quantum Field Theory (QFT), GR turns out to be a non-renormalizable theory, meaning that quantum corrections generate an infinite tower of counterterms that are not present in the original EH action and whose coefficients have to be fixed at every order of perturbation theory \cite{tHooft:1974toh,Goroff:1985th}. From a phenomenological perspective, the $\Lambda$CDM model assumes GR as the underlying gravitational theory, which by itself is not enough to explain, for instance, why the universe appears to be so flat, isotropic, homogeneous and nearly scale invariant. With the inclusion of the cosmic inflation mechanism \cite{Guth:1980zm, Linde:1981mu, Albrecht:1982wi} this can be understood, but the right inflationary model is not yet determined.

%re is in principle a big landscape of inflationary models.

%The pursuit of overcoming these challenges 
The search of the correct inflationary model has motivated the study of gravitational theories beyond GR %along with the standard cosmological model, and great efforts have been made to explore different modifications of GR together with its implications 
\cite{Capozziello:2011et,Clifton:2011jh}.  Natural candidates, arising from both theoretical and observational arguments, are higher-derivative theories of gravity.

The inclusion of generally covariant higher-derivative terms in the gravitational action are necessary to achieve a renormalizable theory \cite{Stelle:1976gc, Stelle:1977ry} and provide a ultraviolet (UV) completion of GR. Furthermore, if one considers the coupling of GR to ordinary quantum matter, these terms become indeed unavoidable \cite{Utiyama:1962sn, Birrell:1982ix}. In the Effective Field Theory (EFT) framework \cite{Donoghue:1994dn,Burgess:2003jk}, Einstein’s gravity can be regarded as an effective low-energy theory that requires higher-derivative corrections that become important as the energy scale increases, such as near black hole singularities or in the early universe. This motivates the study of this kind of theories per se, as any UV completion of gravity has to reduce to them in the IR limit. 

From a phenomenological viewpoint, quadratic gravity is an example of a higher-derivative gravitational theory providing an outstanding cosmological model to describe the inflationary era of the primordial Universe, the Starobinsky model \cite{Starobinsky:1980te}, which does not require extra degrees of freedom and is still one of the favorite candidates for describing the physics of cosmic inflation \cite{Planck:2018jri}.

Nonetheless, higher-derivative theories of gravity are not free of pathologies. As Ostrogradsky proved, theories with higher derivative local terms suffer from instabilities because the corresponding Hamiltonian is unbounded  below \cite{Woodard:2015zca}. This instability usually translates at the quantum level as additional ghosts degrees of freedom, which have a wrong sign kinetic term \cite{Pais:1950za, Asorey:1996hz, Sbisa:2014pzo}, apparently leading to negative norm states and unitarity loss. The different mechanisms to deal with ghosts states are still a puzzling question in the community, although some progress has been made in the recent years \cite{Lee:1969fy,Tomboulis:1983sw,Hawking:2001yt,Bender:2007wu,Biswas:2011ar,Salvio:2018crh,Anselmi:2018kgz,Donoghue:2019fcb,Deffayet:2021nnt,Asorey:2024mkb}. 

The aim of this paper is to analyze the cosmological implications of higher-derivative theories even beyond the popular quadratic gravity, focusing mainly on the sixth-order higher-derivative gravity, which is not only a renormalizable theory but also a super-renormalizable quantum gravity theory, pointing out its fundamental differences with GR and quadratic gravity. The gravitational action considered throughout this paper is
\begin{equation}
    S=\int d^4x\sqrt{-g}\left( \omega_\kappa R+\omega_\Lambda+\theta_R R^2+\theta_C C^{\mu\nu\rho\sigma}C_{\mu\nu\rho\sigma}+\omega_R R\Box R+\omega_C C^{\mu\nu\rho\sigma}\Box C_{\mu\nu\rho\sigma} \right)\,,
    \label{eq:S6}
\end{equation}
where $C_{\mu\nu\rho\sigma}$ denotes the Weyl tensor, $R$ is the Ricci scalar, $\Box \equiv g^{\mu\nu}\nabla_\mu\nabla_\nu$ represents the covariant d'Alembert operator, and $\omega_i, \theta_i$ are the respective coupling parameters governing the higher-order terms in the action.

The structure of this paper is as follows: In Sec.~\ref{sec:cosmological_solutions}, we explore the space of homogeneous and isotropic cosmological solutions of six dimensional theories. Specially, we analyze  how these higher-derivative terms circumvent classical singularity theorems,  yielding singularity-free bouncing scenarios. In Sec.~\ref{sec:tensor_perturbations}, we analyze the propagation of tensor perturbations across various standard cosmological backgrounds and derive the necessary constraints required to guarantee linear stability. Sec.~\ref{sec:scalar_perturbations} is devoted to the investigation of the scalar perturbation sector. Finally, in Sec.~\ref{sec:EFT_observables}, we implement the EFT reduction mechanism to analyze how these higher-order corrections modify the primordial inflationary observables, focusing on the tensor-to-scalar ratio $r$ and the scalar spectral index $n_s$.

\section{Singularity-free solutions}
\label{sec:cosmological_solutions}
Let us start by analyzing the equivalence between some higher-derivative gravity theories formulated in the Jordan frame and General Relativity coupled to matter fields in the Einstein frame via a Weyl transformation \cite{Hindawi:1995an,Gottlober:1989ww}. Following the formalism of \cite{Kuntz:2019lzq}, we show how this correspondence makes possible to evade the classical Hawking-Penrose singularity theorems. In particular, for this section we consider a theory with up to six derivatives in the metric defined by the action
\begin{equation}
    S = \int d^4x \sqrt{-\bar{g}} \, \omega_\kappa \left[ \bar{R} + \frac{\theta_R}{\omega_\kappa}\bar{R}^2 + \frac{\omega_R}{\omega_\kappa} \bar{R}\bar{\Box} \bar{R} \right] \,,
    \label{eq:S6Escalar}
\end{equation}
where terms depending on the Weyl tensor are neglected because our analysis focuses on the evolution of homogeneous and isotropic universes, a geometry in which such terms vanish.

The action \eqref{eq:S6Escalar} is given in the Jordan frame. To establish the equivalence with the Einstein frame, we introduce the following auxiliary scalar variables defined via the functional derivatives of the gravitational Lagrangian,
\begin{equation}
    \phi \equiv \dfrac{\delta \mathcal L}{\delta \bar{R}} = 1 + 2\dfrac{\theta_{R}}{\omega_\kappa}\bar{R} + \dfrac{\omega_R}{\omega_\kappa} \bar{\Box} \bar{R} \,, \quad \text{and} \quad \psi \equiv \dfrac{\delta \mathcal L}{\delta (\bar{\Box} \bar{R})} =\dfrac{\omega_R}{\omega_\kappa} \bar{R} \,.
\end{equation}
By performing a Legendre transformation and integrating the kinetic term by parts, we cast the action into an equivalent scalar-tensor form
\begin{equation}
    S = \int d^4x \sqrt{-\bar{g}} \, \omega_\kappa \left[ \phi \bar{R} - \frac{\omega_\kappa}{\omega_R}(\bar{\nabla} \psi)^2 - \frac{\omega_\kappa}{\omega_R}\psi \left( \phi - 1 - \frac{\theta_R}{\omega_R}\psi \right) \right] \,,
\end{equation}
where the equations of motion for $\phi$ and $\psi$ readily reproduce the defining field constraints.

Next, we perform a Weyl transformation 
\begin{equation}
    g_{\mu\nu} = e^{\chi}\bar{g}_{\mu\nu} \,,  \quad \chi=\ln{\phi} \,,
\end{equation}
 from the Jordan frame to the Einstein frame.
Under this change of variables, and after absorbing total derivatives, the action reduces to General Relativity coupled to two interacting scalar fields,
\begin{equation}
    S = \int d^4x \sqrt{-g} \, \omega_\kappa \left[ R - \frac{3}{2}(\nabla\chi)^2 - \frac{\omega_\kappa}{\omega_R}e^{-\chi}(\nabla \psi)^2 - V(\chi,\psi) \right]\,,
    \label{eq:AccionEinsteinFrame}
\end{equation}
where the interaction potential is defined as
\begin{equation}
    V(\chi,\psi) = \frac{\omega_\kappa}{\omega_R}e^{-2\chi}\psi \left( e^{\chi} - 1 - \frac{\theta_R}{\omega_R}\psi \right) \,.
\end{equation}
Notably, this conformal mapping is well-defined provided that $\phi > 0$, a requirement that translates directly to the geometric condition
\begin{equation}
    1 + 2\frac{\theta_R}{\omega_\kappa}\bar{R} + \frac{\omega_R}{\omega_\kappa} \bar{\Box} \bar{R} > 0 \,.
\end{equation}
The variation of the the action \eqref{eq:AccionEinsteinFrame} with respect to the metric yields the Einstein field equations
\begin{equation}
    R_{\mu\nu} - \frac{1}{2}g_{\mu\nu}R = \frac{3}{2}\left[ (\nabla_\mu \chi)(\nabla_\nu \chi) - \frac{1}{2}g_{\mu\nu}(\nabla\chi)^2 \right] + \frac{\omega_\kappa}{\omega_R}e^{-\chi}\left[ (\nabla_\mu \psi)(\nabla_\nu \psi) - \frac{1}{2}g_{\mu\nu}(\nabla\psi)^2 \right] - \frac{1}{2}g_{\mu\nu}V(\chi,\psi) \,.
    \label{eq:EOMSin}
\end{equation}
By taking the trace of the previous expression, one can eliminate the Ricci scalar and then the Ricci tensor takes the simplified form
\begin{equation}
    R_{\mu\nu} = \frac{3}{2}(\nabla_\mu\chi)(\nabla_\nu\chi) + \frac{\omega_\kappa}{\omega_R}e^{-\chi}(\nabla_\mu\psi)(\nabla_\nu\psi) + \frac{1}{2}g_{\mu\nu}V(\chi,\psi) \,.
\end{equation}

To evaluate the conditions for singularity avoidance, we construct the geometric contraction $\Delta \equiv R_{\mu\nu}k^\mu k^\nu$, where $k^\mu$ is an arbitrary timelike unit vector. The sign of $\Delta$ determines the focusing of geodesics via the Raychaudhuri equation, and it takes the form
\begin{equation}
    \Delta = \frac{3}{2}\dot{\chi}^2 + \frac{\omega_\kappa}{\omega_R}e^{-\chi}\dot{\psi}^2 - \frac{1}{2}V(\chi,\psi) \,,
    \label{eq:Discriminante}
\end{equation}
where the dot denotes differentiation along the timelike congruence defined by $k^\mu$. 

Unlike in standard General Relativity with conventional matter distributions, $\Delta$ can acquire negative values due to the opposite contributions coming from the interaction potential or the kinetic term of $\psi$, depending on the parameter space of the theory. This signature inversion indicates that the classical Hawking-Penrose singularity theorems can be entirely bypassed. Such geometric regularizations are a known feature of higher-derivative gravity, that arise when considering quadratic curvature both as an  effective field theories \cite{Kuipers:2019qby} or a generalized $f(R)$ theories \cite{Alani:2016ail}.

Finally, we linearize the theory around the vacuum by considering small fluctuations of the scalar fields. The quadratic kinetic and mass terms can be simultaneously diagonalized and canonically normalized via the field rotation and rescaling
\begin{equation}
    \tilde{\psi} = \sqrt{\frac{2\omega_\kappa^2}{\omega_R}}\psi \cos\theta - \sqrt{3\omega_\kappa}\chi\sin\theta \,, \quad \text{and} \quad
    \tilde{\chi} = \sqrt{\frac{2\omega_\kappa^2}{\omega_R}}\psi \sin\theta + \sqrt{3\omega_\kappa}\chi\cos\theta \,,
\end{equation}
where the mixing angle is given by $\tan(2\theta) = \sqrt{\frac{2\omega_R\omega_\kappa}{3\theta_R^2}}$. Under this linear transformation, the quadratic Lagrangian density assumes the diagonal form
\begin{equation}
    \mathcal{L}^{(2)} = -\frac{1}{2}(\nabla\tilde{\chi})^2 - \frac{1}{2}(\nabla\tilde{\psi})^2 - \frac{1}{2}m_1^2\tilde{\psi}^2 - \frac{1}{2}m_2^2\tilde{\chi}^2 \,,
\end{equation}
where the mass eigenvalues are given by
\begin{equation}
    m_{1,2}^2 = \frac{-\theta_R}{2\omega_R} \pm \sqrt{\frac{\theta_R^2}{4\omega_R^2} + \frac{\omega_\kappa}{6\omega_R}} \,.
    \label{eq:MasasEscalares}
\end{equation}

To ensure that both field excitations have stable, non-tachyonic masses, we must request that the product of their squared masses is strictly positive, $m_1^2 m_2^2 > 0$. Using the explicit eigenvalues from \eqref{eq:MasasEscalares}, this constraint reduces to
\begin{equation}
    m_1^2 m_2^2 = - \frac{\omega_\kappa}{6\omega_R} > 0 \,.
    \label{eq:condicion_masas}
\end{equation}
This inequality requires that the coupling parameters $\omega_R$ and $\omega_\kappa$ carry opposite signs. Crucially, tracking this sign requirement back into the full Einstein-frame action \eqref{eq:AccionEinsteinFrame} reveals that it flips the sign of the kinetic term for $\psi$ relative to its canonical configuration. 

Consequently, avoiding tachyonic instabilities inevitably forces one of the scalar fields to behave as a ghost. It is precisely this ghost degree of freedom that sources the negative contribution to the geometric contraction $\Delta$ in \eqref{eq:Discriminante}, providing the necessary energy-condition violation to bypass classical spacetime singularities.

We now proceed to study the presence of singularity-free cosmological solutions. To this end, we first derive the generalized Friedmann equations for this theory. Any homogeneous and isotropic solution of the field equations is described by the Friedmann-Lemaître-Robertson-Walker (FLRW) metric
\begin{equation}
    ds^2 = -b(t)^2dt^2 + a(t)^2\left[ \frac{dr^2}{1-\mathcal{K}\,r^2} + r^2d\theta^2 + r^2\sin^2\theta d\phi^2 \right] \,,
\end{equation}
where $a(t)$ is the scale factor dictating the cosmic expansion dynamics and $b(t)$ is the lapse function. General Relativity, the analysis is restricted to the case of a spatially flat universe by imposing $\mathcal{K}=0$.

Substituting this geometric ansatz into the gravitational action \eqref{eq:S6Escalar} yields an effective action $S = \int dt \, \mathcal{L} + S_m$, where the resulting effective Lagrangian depends on higher-order time derivatives of both $a(t)$ and $b(t)$.  In accordance with  Palais' theorem \cite{Palais:1979rca}, the  equations of motion can be obtained directly by varying this simplified symmetry-reduced action via the generalized Euler-Lagrange equations, which read
\begin{equation}
    \frac{\partial\mathcal{L}}{\partial b} - \frac{d}{dt}\left( \frac{\partial \mathcal{L}}{\partial \dot{b} } \right) + \frac{d^2}{dt^2}\left( \frac{\partial \mathcal{L}}{\partial \ddot{b}} \right) - \frac{d^3}{dt^3}\left( \frac{\partial \mathcal{L}}{\partial b^{(III)}} \right) = a^3\rho \,,
    \label{eq:EOMb}
\end{equation}
and
\begin{equation}
    \frac{\partial\mathcal{L}}{\partial a} - \frac{d}{dt}\left( \frac{\partial \mathcal{L}}{\partial \dot{a}} \right) + \frac{d^2}{dt^2}\left( \frac{\partial \mathcal{L}}{\partial \ddot{a}} \right) - \frac{d^3}{dt^3}\left( \frac{\partial \mathcal{L}}{\partial \dddot{a}} \right) + \frac{d^4}{dt^4}\left( \frac{\partial \mathcal{L}}{\partial a^{(IV)}} \right) = -3a^2b p \,,
    \label{eq:EOMa}
\end{equation}
where $\rho$ and $p$ correspond to the energy density and pressure of the cosmic matter source.

Upon evaluating these variations and fixing the standard cosmic time gauge $b(t)=1$, the generalized Friedmann equations for the higher-derivative theory \eqref{eq:S6Escalar} take the form
\begin{equation}
    \begin{aligned}
        & \omega_\Lambda a^3 - \frac{144 \omega_R \dot{a}^6}{a^3} - \frac{936 \omega_R \dot{a}^4 \ddot{a}}{a^2} + \frac{36 \dot{a}^2 \left( -3 \theta_R \dot{a}^2 + 13 \omega_R \ddot{a}^2 + 26 \omega_R \dot{a} \dddot{a} \right)}{a} + \\ 
        &\hspace{1.7cm}+ 72 \left( \omega_R \ddot{a}^3 - 3 \omega_R \dot{a} \ddot{a} \dddot{a} + \dot{a}^2 \left( \theta_R \ddot{a} - 2 \omega_R a^{(IV)} \right) \right) + \\
        &+ 6 a \bigg( \omega_\kappa \dot{a}^2 - 6 \left( \theta_R \ddot{a}^2 + \omega_R \dddot{a}^2 - 2 \omega_R \ddot{a} a^{(IV)} \right) + 12 \dot{a} \left( \theta_R \dddot{a} - \omega_R a^{(V)} \right) \bigg)=a^3\rho \,,
    \end{aligned}
\end{equation}
and
\begin{equation}
    \begin{aligned}
       & \frac{1}{a^4} \Bigg(  3 \bigg( \omega_\Lambda a^6 + 144 \omega_R \dot{a}^6 + 336 \omega_R a \dot{a}^4 \ddot{a}+ 12 a^2 \dot{a}^2 \Big( 3 \theta_R \dot{a}^2 - 13 \omega_R \big( 9 \ddot{a}^2 + 4 \dot{a} \dddot{a} \big) \Big) \bigg) - \\
        & -24 a^3 \bigg( 6 \theta_R \dot{a}^2 \ddot{a} - 13 \omega_R \Big( \ddot{a}^3 + 4 \dot{a} \ddot{a} \dddot{a} + \dot{a}^2 a^{(IV)} \Big) \bigg) +  2 a^4 \bigg( \omega_\kappa \dot{a}^2 + 6 \Big( 3 \theta_R \ddot{a}^2 - 7 \omega_R \dddot{a}^2 - 12 \omega_R \ddot{a} a^{(IV)}\Big)+ \\
        & + 12 \dot{a} \Big( 2 \theta_R \dddot{a} - 3 \omega_R a^{(V)}\Big) \bigg) +  4 a^5 \bigg( \omega_\kappa \ddot{a} + 6 \theta_R a^{(IV)} - 6 \omega_R a^{(VI)}   \bigg) \Bigg)=-3a^2 p \,.
    \end{aligned}
\end{equation}

Despite the algebraic complexity of the generalized field equations, we can systematically analyze the spectrum of valid cosmological solutions. A fundamental result in higher-derivative gravity establishes that vacuum solutions of General Relativity remain also exact solutions for any higher-derivative theory whose action depends exclusively on the Ricci scalar $R$ and the Ricci tensor $R_{\mu\nu}$. In the absence of a cosmological constant, the standard field equations in vacuum imply a Ricci-flat metric ($R_{\mu\nu}=0$). Since the variational terms derived from higher-order modifications are inherently proportional to $R_{\mu\nu}$ or its covariant derivatives, a Ricci-flat geometry implies that these corrections  do vanish, thereby satisfying the full field equations \cite{Li:2015bqa}. We note that this property is not generically preserved if the action explicitly includes the Riemann tensor $R_{\alpha\beta\gamma\delta}$ or the Weyl tensor $C_{\alpha\beta\gamma\delta}$, unless they enter via specific topological combinations such as the Gauss-Bonnet invariant. In four dimensions, the Gauss-Bonnet term constitutes a total derivative and does not contribute to the local dynamics, leaving the vacuum solutions of General Relativity unaltered.

For the specific action \eqref{eq:S6} considered in this work, the choice of an FLRW background implies a conformally flat spacetime where the Weyl tensor vanishes identically. Consequently, any vacuum solution of the standard Friedmann equations constitutes an exact vacuum solution for this sixth-order theory as well.

It is particularly instructive to examine the survival of standard cosmological solutions in the pure higher-derivative limit where the Einstein-Hilbert term is absent ($\omega_\kappa=0$). When evaluating a radiation-dominated power-law expansion ($a(t) \propto t^{1/2}$) or a de Sitter background ($a(t) \propto e^{Ht}$), distinct geometric features emerge. For the radiation-like solution, the Ricci scalar vanishes ($R=0$), whereas the de Sitter metric yields a Ricci tensor proportional to the metric ($R_{\mu\nu} \propto g_{\mu\nu}$). These algebraic properties guarantee that both geometries remain exact vacuum solutions for the sector of the theory dominated by higher-order terms.

Accordingly, any FLRW metric satisfying these geometric constraints that solves General Relativity will automatically satisfy our generalized equations of motion, as shown in \cite{Barrow:2007pm} for $f(R)$ gravity theories. From this analysis, we directly extract the existence condition for the de Sitter solution $a(t)=e^{Ht}$, which emerges in close analogy to General Relativity as
\begin{equation}
    H^2 = -\frac{\omega_\Lambda}{6\omega_\kappa} \,.
    \label{eq:condicion_hubble_desitter}
\end{equation}

\begin{figure}[H]
    \centering
    \begin{subfigure}[c]{0.42\textwidth}
        \includegraphics[width=\textwidth]{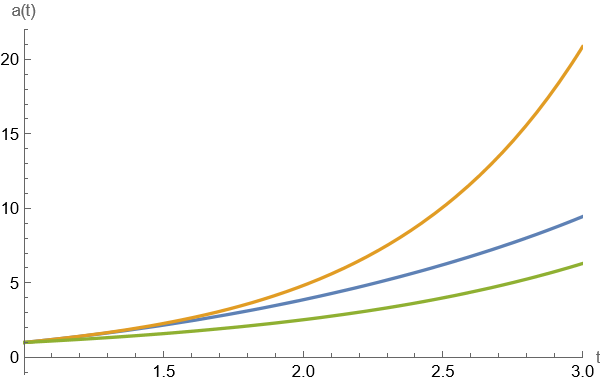}
        \caption{Model parameters $\omega_\Lambda=-100$, $\omega_\kappa=20$, $\theta_R=10$, $\omega_R=-1$, with matter content $\rho_0=10$ and $\omega=0$. The initial conditions are $a(1)=1$, $\dot{a}(1)=1.9$, $\ddot{a}(1)=1.5$, $\dddot{a}(1)=1.5$, and $a^{(IV)}(1)=1.6$.}
        \label{fig:Fig1a}
    \end{subfigure}
    \hfill
    \begin{subfigure}[c]{0.42\textwidth}
        \includegraphics[width=\textwidth]{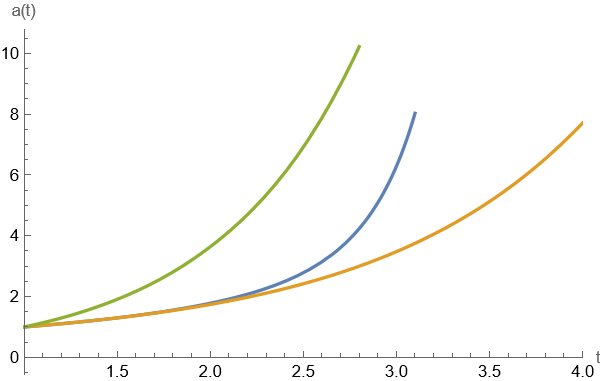}
        \caption{Model parameters $\omega_\Lambda=-200$, $\omega_\kappa=20$, $\theta_R=14$, $\omega_R=1$, in vacuum ($\rho_0=0$). The initial conditions are $a(1)=1$, $\dot{a}(1)=0.5$, $\ddot{a}(1)=0.3$, $\dddot{a}(1)=0.5$, and $a^{(IV)}(1)=0.6$.}
        \label{fig:Fig1b}
    \end{subfigure}
    \hfill
    \begin{subfigure}[t]{0.1\textwidth}
        \includegraphics[width=\textwidth]{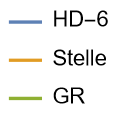}
    \end{subfigure}
    \caption{Numerical solutions displaying expanding universes.}
    \label{fig:Fig1}
\end{figure}

\begin{figure}[H]
    \centering
    \begin{subfigure}[c]{0.42\textwidth}
        \includegraphics[width=\textwidth]{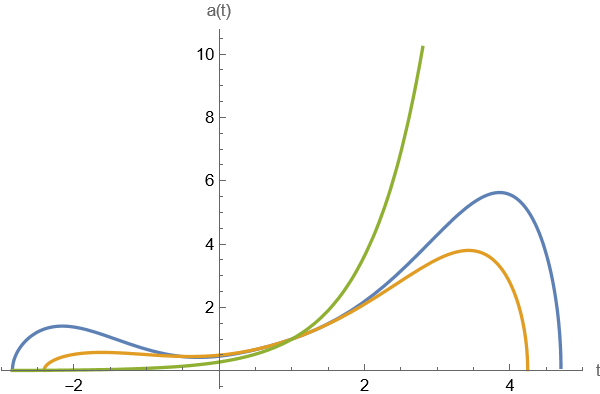}
        \caption{Model parameters $\omega_\Lambda=-200$, $\omega_\kappa=20$, $\theta_R=-5$, and $\omega_R=1$.}
        \label{fig:Fig2a}
    \end{subfigure}
    \hfill
    \begin{subfigure}[c]{0.42\textwidth}
        \includegraphics[width=\textwidth]{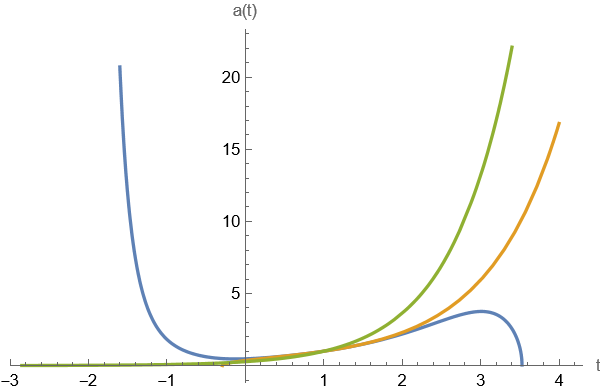}
        \caption{Model parameters $\omega_\Lambda=-200$, $\omega_\kappa=20$, $\theta_R=5$, and $\omega_R=1$.}
        \label{fig:Fig2b}
    \end{subfigure}
     \hfill
    \begin{subfigure}[t]{0.1\textwidth}
        \includegraphics[width=\textwidth]{Figuras/Legend.png}
    \end{subfigure}
    \caption{Numerical solutions in vacuum revealing recollapsing profiles that terminate in a Big Crunch. The chosen initial conditions are $a(1)=1$, $\dot{a}(1)=0.8$, $\ddot{a}(1)=0.6$, $\dddot{a}(1)=0.5$, and $a^{(IV)}(1)=0.6$.}
    \label{fig:Fig2}
\end{figure}

Next, we analyze the spectrum of numerical cosmological solutions obtained within this sixth-order higher-derivative theory. Initially, we identify standard expanding solutions that are dynamically analogous to General Relativity, though featuring modified expansion rates determined by the specific choices of the coupling parameters (\Cref{fig:Fig1}). From now on, in all the plots, the green line corresponds to General Relativity, the orange line corresponds to Stelle fourth order gravity (or equivalently, to the Starobinsky model $f(R)=R+\alpha R^2$ in FLRW backgrounds) and the blue line corresponds to sixth-order higher-derivative gravity.
Furthermore, it should be remarked that the sixth-order gravity yields novel cosmological behaviors that are absent in standard General Relativity. This includes recollapse scenarios where the universe undergoes an initial expansion phase, reaches a maximum turning point, and subsequently enters into a contracting phase that terminates in a Big Crunch singularity (\Cref{fig:Fig2}). Notably, while this recollapsing behavior is strictly restricted to the $\theta_R < 0$ regime in Stelle quadratic gravity, the additional structure of the six-derivative super-renormalizable theory renders this scenario to be viable even when $\theta_R > 0$. These dynamics remain robust under the inclusion of different matter sources.

\begin{figure}[H]
    \centering
    \begin{subfigure}[c]{0.44\textwidth}
        \includegraphics[width=\textwidth]{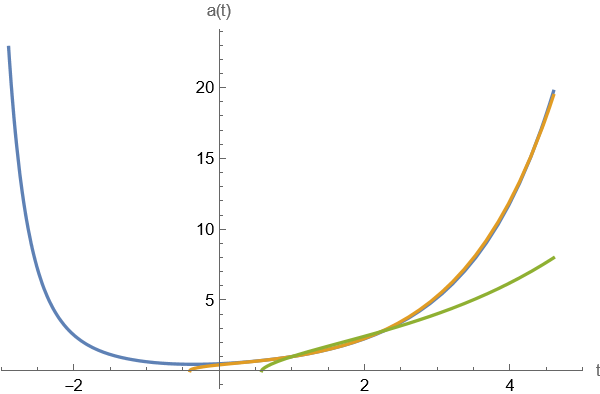}
        \caption{Real pole regime with parameters $\omega_\Lambda=-1$, $\omega_\kappa=1$, $\theta_R=1$, $\omega_R=1$, and matter distribution $\rho_0=15$, $\omega=0$. Initial configurations are set to $a(1)=1$, $\dot{a}(1)=0.8$, $\ddot{a}(1)=0.7$, $\dddot{a}(1)=0.5$, and $a^{(IV)}(1)=0.6$.}
        \label{fig:Fig3a}
    \end{subfigure}
    \hfill
    \begin{subfigure}[c]{0.44\textwidth}
        \includegraphics[width=\textwidth]{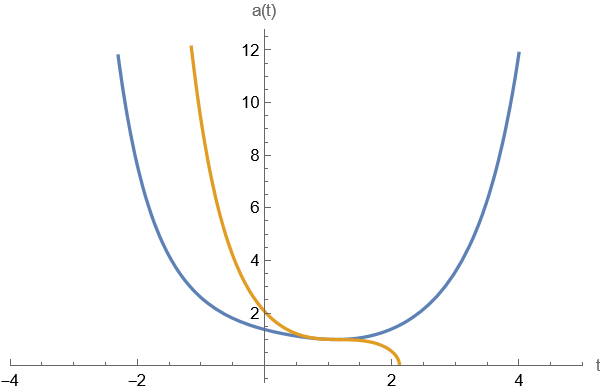}
        \caption{Complex pole regime with parameters $\omega_\Lambda=10$, $\omega_\kappa=10$, $\theta_R=-1$, $\omega_R=-1$, in vacuum ($\rho_0=0$). Initial configurations are set to $a(1)=1$, $\dot{a}(1)=-0.1$, $\ddot{a}(1)=0.7$, $\dddot{a}(1)=0.8$, and $a^{(IV)}(1)=0.9$.}
        \label{fig:Fig3b}
    \end{subfigure}
     \hfill
    \begin{subfigure}[t]{0.1\textwidth}
        \includegraphics[width=\textwidth]{Figuras/Legend.png}
    \end{subfigure}
    \caption{Singularity-free numerical solutions representing the real pole spectrum (left plot) and the complex pole spectrum (right plot).}
    \label{fig:Fig3}
\end{figure}

Finally, we explore the parameter space allowing for fully singularity-free cosmological solutions (\Cref{fig:Fig3}). In Stelle gravity, regular solutions are restricted exclusively to the negative parameter regime $\theta_R < 0$ when coupling to matter \cite{Asorey:2024oxw}. In contrast, the six-derivative theory allows for non-singular configurations even when $\theta_R > 0$. This singularity avoidance holds for configurations possessing both real poles (\Cref{fig:Fig3a}) and complex poles (\Cref{fig:Fig3b}), albeit with a notable distinction, models with real poles yield singularity-free evolutions irrespective of the matter content, whereas models in the complex poles regime require a pure vacuum configuration with a positive cosmological constant $\omega_\Lambda > 0$.
Notice that in this case there is no GR solution, since with our convention it requires $\omega_\Lambda < 0$.

\section{Tensor perturbations}
\label{sec:tensor_perturbations}
In this section we analyze the propagation of tensor perturbations over cosmological backgrounds in the context of the sixth-order gravity theory defined by the action \eqref{eq:S6}. This approach parallels the methodology used in \cite{Fabris:2011qq,Shapiro:2014fsa,Peter:2017xxf}, which incorporated quantum corrections due to the trace anomaly. In this work, such corrections are neglected to focus strictly on the purely classical dynamics of the Lagrangian.

We consider a FLRW background metric with transverse and traceless spin-2 tensor perturbations $h_{ij}(\mathbf{x}, \eta)$, parameterized by conformal time $\eta$, such that $\partial_i h^{ij}=0$ and $h^i_i=0$, yielding the following line element
\begin{equation}
    ds^2 = a^2(\eta) \left[ -d\eta^2 + (\delta_{ij} + h_{ij})dx^i dx^j \right] \,.
    \label{eq:metrica_perturbada}
\end{equation}

Linearized equations of motion for tensor modes are obtained by expanding the gravitational action up to second order in $h_{ij}$. By collecting the perturbative contributions from each term of the action, the second-order variation $\delta^{(2)}\mathcal{L}$ leads to an effective  Lagrangian quadratic in tensorial perturbations which is governed  
 by spatial Laplacian operators $\Delta$ alongside conformal time derivatives up to fourth order.

The matter sector we note that the unperturbed background energy-momentum tensor takes the standard perfect fluid form $T^{(0)00} = \rho/a^2$ and $T^{(0)ij} = (P/a^2)\delta^{ij}$. Since the background equations of motion are satisfied at zeroth order ($E^{(0)\mu\nu} = T^{(0)\mu\nu}$), the first-order variation of the field equations directly couples to the perturbed spatial stresses
\begin{equation}
    \delta^{(1)} E^{ij} = \delta^{(1)} T^{ij} = -\frac{P(\eta)}{a^2}h^{ij} \,.
    \label{eq:PertEOM}
\end{equation}

To relate this  variation  of the macroscopic energy-momentum to the perturbative expansion of the Lagrangian, we use the standard variational identity connecting the second variation of the action to the linearized equations of motion. Since the first variation of the metric determinant vanishes identically for transverse traceless perturbations ($\delta^{(1)}\sqrt{-g} = 0$), the quadratic action functional reduces to
\begin{equation}
    \delta^{(2)}S = \frac{1}{2}\int d^4x \,  \int d^4y \, \delta g_{\mu\nu}(x) \frac{\delta^2S}{\delta g_{\mu\nu}(x)\delta g_{\alpha\beta}(y)}\delta g_{\alpha\beta}(y)  = -\frac{1}{2}\int d^4x \sqrt{-g} \, \delta^{(1)}E^{\mu\nu} \delta g_{\mu\nu} \,.
\end{equation}
This relation confirms that extracting the functional derivative of the second-order action with respect to the perturbation $h_{ij}$ readily yields the exact linearized field equations governing the propagation of the tensor modes over the cosmological background.
If expressed in terms of the spatial perturbations \( h_{ij} \) and using \eqref{eq:PertEOM} for the expression of \( \delta^{(1)}E^{\mu\nu} \), we finally obtain the relation,
\begin{equation}
    \frac{\delta^2S}{\delta h_{ij}h_{kl}}h_{kl}=a^4P(\eta)h^{ij}\,.
\end{equation}

Applying this variational identity yields to the full linearized equations of motion for the tensor perturbations
\begin{equation}
    \begin{aligned}
        &\omega_Ch^{(VI)}_{ij}-6\omega_C\mathcal{H} h^{(V)}_{ij}+\left[2\omega_C\left( 3\mathcal{H}^2-4\mathcal{H}^{'} \right) -a^2\theta_C \right]h^{(IV)}_{ij}+2\omega_C\left( 8\mathcal{H}^3 - 2\mathcal{H}\mathcal{H}' - 3\mathcal{H}''\right) h^{'''}_{ij}+\\
        &+\Big[\omega_C \left( -2\mathcal{H}''' - 4\mathcal{H}\mathcal{H}'' - 8(\mathcal{H}')^2 + 52\mathcal{H}^2\mathcal{H}' - 24\mathcal{H}^4\right)+\\
        &\hspace{0.65cm}+\omega_R \left(-6\mathcal{H}''' + 36\mathcal{H}^2\mathcal{H}' \right) +\frac{\omega_{\kappa}}{2}a^4+6\theta_{R}a^2\left( \mathcal{H}^2+\mathcal{H}' \right) \Big]h''_{ij} +\\
        &+\left[ 6\omega_R\left( -\mathcal{H}^{(4)} + 2\mathcal{H}\mathcal{H}''' + 6\mathcal{H}^2\mathcal{H}'' + 12\mathcal{H}(\mathcal{H}')^2 - 12\mathcal{H}^3\mathcal{H}' \right)+ \omega_{\kappa} a^4\mathcal{H}+6a^2\theta_{R}\left( 2\mathcal{H}\mathcal{H}'+\mathcal{H}'' \right) \right]h'_{ij}-\\
        &-3\omega_C\Delta h_{ij}^{(IV)}+12\omega_C\mathcal{H}\Delta h_{ij}^{'''}+\left[ 4\omega_C\left( 4\mathcal{H}^{'}-3\mathcal{H}^2 \right)+42\omega_R\left( \mathcal{H}'+\mathcal{H}^2 \right) +2a^2\theta_C \right]\Delta h''_{ij} +3\omega_C \Delta^2 h''_{ij}+\\
        & +\left[ 10\omega_C\left( \mathcal{H}^{''}-2\mathcal{H}\mathcal{H}^{'} \right)+42\omega_R\left( \mathcal{H}^{''}-2\mathcal{H}^{3} \right) \right]\Delta h'_{ij}-6\omega_C\mathcal{H}\Delta^2h'_{ij}+\\
        &+\Big[-\omega_C\left(  2\mathcal{H}''' + 4\mathcal{H}\mathcal{H}'' + 8(\mathcal{H}')^2 - 52\mathcal{H}^2\mathcal{H}' + 24\mathcal{H}^4 \right)- \frac{\omega_{\kappa}}{2}a^4-6a^2\theta_{R}\left( \mathcal{H}^2+\mathcal{H}' \right)-\\
        &\hspace{0.7cm}-\omega_R\left(  -27\mathcal{H}''' + 42\mathcal{H}\mathcal{H}'' + 162\mathcal{H}^2\mathcal{H}' - 84\mathcal{H}^4 + 42(\mathcal{H}' + \mathcal{H}^2) \right)  \Big]\Delta h_{ij}+\\
        &+\left[\omega_C(6\mathcal{H}^2 - 8\mathcal{H}')-a^2\theta_C\right]\Delta^2 h_{ij}+\omega_C\nabla^3 h_{ij}+\\
        &+\Bigg[ \frac{\omega_{\Lambda}}{2} a^6+\omega_{\kappa} a^4(\mathcal{H}^2+2\mathcal{H}')+6a^2\theta_R\left( 2\mathcal{H}''' - 2\mathcal{H}\mathcal{H}'' + (\mathcal{H}')^2 - 12\mathcal{H}^2\mathcal{H}' + 3\mathcal{H}^4  \right)+ \\
        &+ \omega_R\big(  -12\mathcal{H}^{(V)} + 84\mathcal{H}\mathcal{H}^{(IV)} + 60\mathcal{H}'\mathcal{H}''' - 48\mathcal{H}^2\mathcal{H}''' + 18(\mathcal{H}'')^2 + 552\mathcal{H}\mathcal{H}'\mathcal{H}'' - 228\mathcal{H}^3\mathcal{H}'' + \\
        &+ 282(\mathcal{H}')^3 + 828\mathcal{H}^2(\mathcal{H}')^2 - 360\mathcal{H}^4\mathcal{H}'  + 72\mathcal{H}^6 \big) \Bigg]h_{ij}=-P(\eta)a^6h_{ij}\,,
    \end{aligned}
    \label{eq:PertEOM6}
\end{equation}
where $\mathcal H = a'/a$ is the conformal Hubble parameter and the prime denotes the derivative with respect to conformal time. Notice that the coefficient accompanying the undifferentiated $h_{ij}$ term corresponds precisely to the spatial component of the background Friedmann equations. Consequently, when evaluating these perturbations on a background that exactly satisfies the equations of motion, this particular term identically vanishes on-shell.

We analyze now the evolution of the tensor perturbations across different cosmological backgrounds. 
Given the spatial homogeneity of the background, we apply a standard Fourier mode decomposition to the perturbation field,
\begin{equation}
    h_{ij}(\eta,\mathbf{x}) = \sum_{\lambda=+,\times} \int \frac{d^3\mathbf{k}}{(2\pi)^3} h^{(\lambda)}_{ij}(\eta,\mathbf{k}) e^{i\mathbf{k}\cdot\mathbf{x}} \, .
    \label{eq:fourier_expansion}
\end{equation}
Thanks to the linearity of the governing perturbation equations, each Fourier mode $\mathbf{k}$ evolves independently. 
We can therefore factorize the tensorial structure as $h^{(\lambda)}_{ij}(\eta,\mathbf{k}) = h^{(\lambda)}_{\mathbf{k}}(\eta) e_{ij}^{(\lambda)}(\hat{\mathbf{k}})$, where $e_{ij}^{(\lambda)}(\hat{\mathbf{k}})$ denotes the standard transverse and traceless polarization basis tensors.

The subsequent dynamics of these fluctuations are determined by their initial conditions at the reference conformal time $\eta_i$. 
Assuming a scale-dependent initialization governed by the wavenumber $k$, the mode amplitudes and their corresponding conformal time derivatives satisfy
\begin{equation}
    h_k(\eta_{i}) = \frac{1}{\sqrt{2k}} \, , \quad 
    h'_k(\eta_{i}) = \sqrt{\frac{k}{2}} \, , \quad \dots \, , \quad
    h^{(n)}_k(\eta_{i}) = \frac{k^{(2n-1)/2}}{\sqrt{2}} \, ,
    \label{eq:condiciones_iniciales}
\end{equation}
where $h^{(n)}_k$ represents the $n$-th order derivative with respect to conformal time.
\subsection{Perturbations on a Minkowski background}
We study the equations of motion for tensor perturbations on a flat background for two scenarios: the quadratic gravity regime (Stelle gravity, where $\omega_R=\omega_C=0$) and the six derivatives theory.

\subsubsection{Quadratic gravity}
For a flat background where $a(\eta)=1$ (implying $t=\eta$), the equations of motion simplify to
\begin{equation}
    -\theta_C h_{\mathbf{k}}^{(IV)}(t) + \left(\frac{\omega_{\kappa}}{2}-2\theta_C k^2\right)h_{\mathbf{k}}''(t) + k^2\left( \frac{\omega_{\kappa}}{2}-\theta_C k^2\right)h_{\mathbf{k}}(t)=0\,.
    \label{eq:eom_minkowski}
\end{equation}
The analytical solution to this differential equation, satisfying the initial conditions at $t=0$, is given by
\begin{equation}
    h_{\mathbf{k}}(t) = \frac{\omega_{\kappa}-4\theta_C k^2}{\sqrt{2k}\,\omega_{\kappa}}\left[ \cos(kt)+\sin(kt) \right] - \frac{\theta_C\sqrt{2k^3}}{\omega_{\kappa} \omega}\left[ (k-\omega)e^{-\omega t}-(k+\omega)e^{\omega t} \right]\,,
    \label{eq:solucion_minkowski}
\end{equation}
where we have defined the characteristic frequency $\omega=\sqrt{-k^2+\frac{\omega_{\kappa}}{2\theta_C}}$. 
The solution diverges exponentially if $\omega$ is real, a condition satisfied when $k^2 < \frac{\omega_{\kappa}}{2\theta_C}$, otherwise the solution remains bounded since it is purely oscillating.

This equation of motion can be rewritten in a compact form in terms of the d'Alembertian operator $\Box$ as
\begin{equation}
    \left(-\theta_C \Box^2 - \frac{\omega_{\kappa}}{2}\Box\right)h_{ij}(x) = -\theta_C\Box\left(\Box + \frac{\omega_{\kappa}}{2\theta_C}\right)h_{ij}(x) = 0\,.
    \label{eq:eom_dalembertian}
\end{equation}
By means of a Fourier transform, we find the propagator associated with the spin-2 sector of the theory,
\begin{equation}
    G(k) = \frac{-1}{\theta_C\left(k^4 - \frac{\omega_{\kappa}}{2\theta_C}k^2\right)} = \frac{1}{\omega_{\kappa}}\left( \frac{1}{k^2} - \frac{1}{k^2+m_2^2} \right)\,,
    \label{eq:propagador_spin2}
\end{equation}
where $m_2^2 = -\frac{\omega_{\kappa}}{2\theta_C}$. The particle spectrum of the theory is strictly determined by the signs of the parameters $\omega_{\kappa}$ and $\theta_C$. Throughout this work, we enforce $\omega_{\kappa} > 0$ to guarantee that the massless graviton mode possesses the correct kinetic sign, ensuring positive energy and preventing ghost instabilities. Consequently, the choice of $\theta_C$ yields two distinct physical regimes. First, the non-tachyonic regime ($\theta_C < 0$, or equivalently $m_2^2 > 0$) describes a theory featuring the standard Einstein-Hilbert massless graviton alongside a massive, non-tachyonic ghost-like degree of freedom. On the other hand, the tachyonic regime ($\theta_C > 0$, implying $m_2^2 < 0$) yields a spectrum wherein the standard massless graviton is instead accompanied by a tachyonic, ghost-like excitation.

\subsubsection{Sixth-order gravity}
On a Minkowski background, the  equation of motion of tensorial perturbations 
for the full six derivatives theory takes the form
\begin{equation}
    2\omega_C h^{(VI)} + 2(3k^2\omega_C-\theta_C)h^{(IV)} + (6k^4\omega_C - 4k^2\theta_C + \omega_\kappa)h'' + k^2(2k^4\omega_C - 2k^2\theta_C + \omega_\kappa)h = 0\,,
    \label{eq:eom_minkowski_6der}
\end{equation}
yielding the general homogeneous solution
\begin{equation}
    h_{\mathbf{k}}(t) = C_1 e^{ikt} + C_2 e^{-ikt} + C_3 e^{\alpha_1 t} + C_4 e^{-\alpha_1 t} + C_5 e^{\alpha_2 t} + C_6 e^{-\alpha_2 t}\,,
    \label{eq:solucion_general_6der}
\end{equation}
where $C_i$ are integration constants and the characteristic exponents are given by
\begin{subequations}
     \label{eq:tensormass6Hd}
    \begin{align}
        \alpha_1^2 &= \frac{\theta_C - 2k^2\omega_C}{2\omega_C} + \frac{1}{2\omega_C}\sqrt{\theta_C^2 - 2\omega_\kappa\omega_C}\,, \\
        \alpha_2^2 &= \frac{\theta_C - 2k^2\omega_C}{2\omega_C} - \frac{1}{2\omega_C}\sqrt{\theta_C^2 - 2\omega_\kappa\omega_C}\,.
    \end{align}
\end{subequations}

To guarantee the classical stability of the theory and ensure that the perturbations remain bounded, the exponents $\alpha_{1,2}$ must be purely imaginary, meaning $\alpha_{1,2}^2$ must be real and negative. This requirement imposes the following simultaneous constraints
\begin{equation}
    \theta_C^2 - 2\omega_\kappa\omega_C > 0\,,\qquad \frac{\theta_C - 2k^2\omega_C}{2\omega_C} < 0\,,\qquad \left| \frac{\theta_C - 2k^2\omega_C}{2\omega_C} \right| > \frac{1}{2\omega_C}\sqrt{\theta_C^2 - 2\omega_\kappa\omega_C} \,.
\end{equation}
The first constraint establishes to avoid exponential divergences requires the poles of the propagator must remain strictly real.

\begin{figure}[htbp]
    \includegraphics[width=0.6\textwidth]{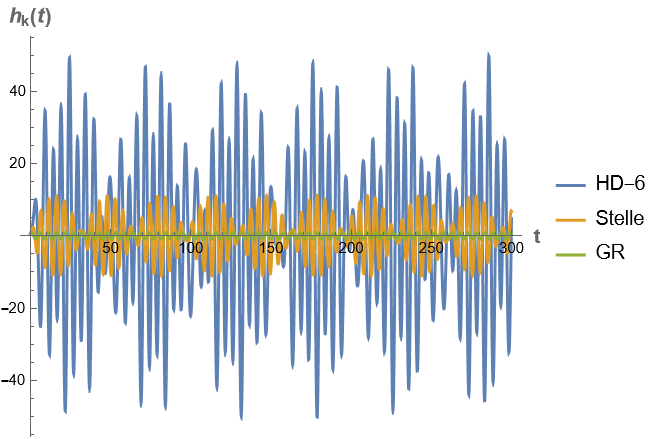}
   \caption{Time evolution of tensor perturbations $h_k(t)$ in a Minkowski background with parameters $k=1.2$, $\omega_\kappa=1$, $\theta_R=1$, $\theta_C=-1$, $\omega_R=1$, $\omega_C=-2$.}
    \label{fig:Fig4}
\end{figure}

\Cref{fig:Fig4} illustrates the numerical time evolution of the perturbations across the analyzed theories. As the maximum derivative order of the theory increases, there is a corresponding growth in the perturbation amplitude fundamentally driven by the higher-order $k^n$ terms arising into the equations of motion.

As before, the equation of motion can be factored using the d'Alembertian operator, yielding
\begin{equation}
    \left( -2\omega_C\Box^3 - 2\theta_C\Box^2 - \omega_\kappa\Box \right)h = 0\,.
    \label{eq:eom_factorized}
\end{equation}
Consequently, the corresponding propagator in Fourier space takes the nested structure
\begin{align}
    G(k) &= \frac{1}{2\omega_C k^6 - 2\theta_C k^4 + \omega_\kappa k^2} \nonumber\\
         &= \frac{1}{2\omega_C} \frac{1}{k^2(k^2+\lambda_1)(k^2+\lambda_2)} \nonumber\\
         &= \frac{1}{2\omega_C} \left[ \frac{1}{\lambda_1\lambda_2}\frac{1}{k^2} - \frac{1}{\lambda_2-\lambda_1}\frac{1}{k^2+\lambda_1} + \frac{1}{\lambda_2-\lambda_1}\frac{1}{k^2+\lambda_2} \right]\,,
    \label{eq:propagador_decomposition}
\end{align}
where the mass parameters $\lambda_{1,2}$ are given by
\begin{subequations}
    \begin{align}
        \lambda_1 &= \frac{-\theta_C}{2\omega_C} - \frac{1}{2\omega_C}\sqrt{\theta_C^2 - 2\omega_C\omega_\kappa}\,, \\
        \lambda_2 &= \frac{-\theta_C}{2\omega_C} + \frac{1}{2\omega_C}\sqrt{\theta_C^2 - 2\omega_C\omega_\kappa}\,.
    \end{align}
\end{subequations}

The tensorial dynamics is thus completely determined by a single massless graviton and two massive excitations, whose physical masses and residues depend uniquely on the underlying Lagrangian couplings. Crucially, the relative signs between these residues are an unavoidable  algebraic consequence of the partial fraction decomposition rather than an arbitrary choice of parameters.
\subsection{Radiation-dominated universe}
\subsubsection{Quadratic gravity}

Let us now consider a radiation-dominated background characterized by the scale factor $a(\eta)\propto\eta$. Because this cosmic evolution constitutes an exact solution to the background Friedmann equations, the corresponding equation of motion for the tensor perturbations reduces to
\begin{equation}
    -2\theta_C h^{(IV)}(\eta) + (\omega_{\kappa}\eta^2-4\theta_C k^2)h''(\eta) + 2\omega_{\kappa}\eta h'(\eta) + k^2(\omega_{\kappa}\eta^2-2k^2\theta_C)h(\eta)=0\,.
    \label{eq:eom_rad_stelle_full}
\end{equation}
Although the stability of past singularities in the $t\to 0$ limit within quadratic gravity frameworks has been extensively analyzed in the literature \cite{Barrow:2007pm,Middleton:2008rh}, the primary focus of this work lies in the late-time behavior of the perturbations, once they enter the horizon. Asymptotically tracking the deep sub-horizon limit ($k\eta \gg 1$), the leading-order contributions to the differential coefficients simplify \eqref{eq:eom_rad_stelle_full} to
\begin{equation}
    -2\theta_C h^{(IV)}(\eta) + \omega_{\kappa}\eta^2 h''(\eta) + 2\omega_{\kappa}\eta h'(\eta) + k^2\omega_{\kappa}\eta^2 h(\eta) = 0\,.
    \label{eq:eom_rad_stelle_asymp}
\end{equation}

To obtain analytical insights into this asymptotic regime, we implement the WKB approximation via the ansatz $h(\eta) \sim e^{S(\eta)}$. Assuming that the higher-order derivatives of the phase function are subleading compared to powers of its first derivative, the characteristic equation takes the form
\begin{equation}
    -2\theta_C (S')^4 + \omega_{\kappa}\eta^2 (S')^2 + 2\omega_{\kappa}\eta S' + k^2\omega_{\kappa}\eta^2 \approx 0\,.
    \label{eq:wkb_characteristic}
\end{equation}
A dominant balance analysis reveals that the equation \eqref{eq:wkb_characteristic} decouples into two distinct operational branches governed by different physical scales.

The first branch yields the high-frequency fast modes, which arise from the dominant balance between the higher-derivative terms. Neglecting the linear and zero-th order terms in $S'$, the characteristic relation simplifies to
\begin{equation}
    -2\theta_C (S')^4 + \omega_{\kappa}\eta^2 (S')^2 \approx 0\,,
    \label{eq:eqMR}
\end{equation}
from which we find the phase derivative $(S')^2 \approx \omega_{\kappa}\eta^2 / (2\theta_C)$. Integrating this expression yields the phase $S(\eta) \approx \pm i \sqrt{-\omega_{\kappa} / (2\theta_C)}\, \eta^2 / 2$, which dictates the fast oscillatory solutions
\begin{equation}
    h_{1,2}(\eta) \sim \exp\left(\pm i \sqrt{\frac{-\omega_{\kappa}}{2\theta_C}} \frac{\eta^2}{2}\right)\,.
\end{equation}
This approximation is self-consistent in the ultraviolet regime, as the substitution of the scaling $S' \sim \mathcal{O}(\eta)$ ensures that the selected terms scale as $\mathcal{O}(\eta^4)$, comfortably dominating the neglected terms which scale only as $\mathcal{O}(\eta^2)$.

The second branch describes the slow modes, representing the standard General Relativity limit. This regime is governed by the balance of the lower-derivative terms under the assumption that the higher-derivative operator $-2\theta_C(S')^4$ is suppressed. The reduced equation
\begin{equation}
    \omega_{\kappa}\eta^2 (S')^2 + 2\omega_{\kappa}\eta S' + k^2\omega_{\kappa}\eta^2 \approx 0
\end{equation}
exactly mimics the standard tensor wave equation in a radiation-dominated universe under classical GR. Solving for the phase derivative yields
\begin{equation}
    S' = \frac{-2\eta \pm \sqrt{4\eta^2 - 4k^2\eta^4}}{2\eta^2} \approx -\frac{1}{\eta} \pm i k\,,
\end{equation}
which integrates directly to $S(\eta) = -\ln(\eta) \pm i k\eta$. This provides the long-wavelength solutions
\begin{equation}
    h_{3,4}(\eta) = \frac{1}{\eta} e^{\pm i k\eta}\,.
\end{equation}
Since $S' \sim \mathcal{O}(1)$ for this branch, the fourth-order derivative term scales as $\mathcal{O}(\eta^0)$, which is negligible with respect to the dominant $\mathcal{O}(\eta^2)$ terms.

Combining both branches, the full asymptotic solution for the tensor perturbation is given by the linear combination
\begin{equation}
    h(\eta) \approx C_1 e^{i\Omega \eta^2} + C_2 e^{-i\Omega \eta^2} + \frac{1}{\eta}\left( C_3 e^{ik\eta} + C_4 e^{-ik\eta} \right)\,,
    \label{eq:solucion_asintotica_radiacion}
\end{equation}
where we have defined the characteristic frequency scale $\Omega = \frac{1}{2}\sqrt{-\omega_{\kappa} / (2\theta_C)}$.

The physical viability of the late-time solution is heavily dependent on the relative signs of the structural parameters. If $\theta_C$ and $\omega_{\kappa}$ possess opposite signs, the square root argument remains positive, leading to bounded, high-frequency oscillations. Conversely, if the parameters share the same sign, the phase exponent becomes real, triggering a fatal exponential growth. Because the standard cosmological framework requires $\omega_{\kappa}>0$ to successfully reproduce the low-energy GR limit, the requirement of late-time stability uniquely enforces the constraint $\theta_C < 0$. This stability threshold matches the parameter space required for a tachyon-free spectrum.

\subsubsection{Sixth-order gravity}
In the full sixth-order theory, we obtain  the explicit relation \begin{align}
    &\frac{\omega_C}{\eta^2}h^{(VI)} - 6\frac{\omega_C}{\eta^3}h^{(V)} + \left[ \frac{\omega_C}{\eta^2}\left( \frac{14}{\eta^2}+3k^2-\theta_C \right) \right]h^{(IV)} + \frac{\omega_C}{\eta^3}\left( \frac{8}{\eta^2}-12k^2 \right)h''' \nonumber\\
    &\quad -\left[ 2k^2\theta_C + \frac{\omega_C}{\eta^2}\left( 3k^4+\frac{28k^2}{\eta^2}-\frac{80}{\eta^4} \right) - \frac{\omega_\kappa}{2}\eta^2 \right]h'' \nonumber\\
    &\quad + \left[ \omega_\kappa\eta - \frac{\omega_C k^2}{\eta^3}\left( \frac{40}{\eta^2}+6k^2 \right) \right]h' \nonumber\\
    &\quad + \left[ \frac{\omega_C k^2}{\eta^2}\left( \frac{80}{\eta^4}+\frac{14k^2}{\eta^2}+k^4 \right) - \theta_C k^4 + \frac{\omega_\kappa}{2}\eta^2 k^2 \right]h = 0\,,
    \label{eq:PertEOM6rad}
\end{align}
after substituting the radiation background scale factor into the complete perturbation equation.

Asymptotically tracking the deep sub-horizon limit ($k\eta \gg 1$), we retain only the leading-order contributions within each coefficient. This simplifies equation \eqref{eq:PertEOM6rad} to 
\begin{equation}
    \frac{\omega_C}{\eta^2}h^{(VI)} - 6\frac{\omega_C}{\eta^3}h^{(V)} - \theta_C h^{(IV)} - 12\frac{\omega_C k^2}{\eta^3}h''' + \frac{\omega_\kappa}{2}\eta^2 h'' + \omega_\kappa\eta h' + \frac{\omega_\kappa}{2}\eta^2 k^2 h = 0\,.
    \label{eq:asymptotic_eq}
\end{equation}
The application of the WKB ansatz $h(\eta) \sim e^{S(\eta)}$ yields the characteristic phase equation
\begin{equation}
    \frac{\omega_C}{\eta^2}(S')^6 - 6\frac{\omega_C}{\eta^3}(S')^5 - \theta_C (S')^4 - 12\frac{\omega_C k^2}{\eta^3}(S')^3 + \frac{\omega_\kappa}{2}\eta^2 (S')^2 + \omega_\kappa\eta S' + \frac{\omega_\kappa}{2}\eta^2 k^2 \approx 0\,.
    \label{eq:wkb_characteristic_6th}
\end{equation}
Repeating the dominant balance analysis carried out  for quadratic gravity, we identify two tracking operational modes.

The slow modes match the standard General Relativity tracking behavior described previously, where the lower-derivative operators dominate the characteristic expression. This branch yields the traditional power-law envelope solutions
\begin{equation}
    h(\eta) \approx \frac{1}{\eta} e^{\pm i k\eta}\,.
\end{equation}
A direct consistency check verifies that the GR terms scale as $\mathcal{O}(\eta^2)$, easily overwhelming the higher-derivative terms which scale at or below $\mathcal{O}(\eta^0)$ under this phase profiles.

The fast modes track rapidly varying fluctuations modeled by the high-frequency scaling profile $S'\sim \lambda\eta$. Under this assumption, the dominant balance is given by  the leading $\mathcal{O}(\eta^4)$ operators. Dropping the subleading terms, the algebraic system reduces to the biquadratic relation
\begin{equation}
    \frac{\omega_C}{\eta^2}(\lambda\eta)^6 - \theta_C(\lambda\eta)^4 + \frac{\omega_\kappa}{2}\eta^2(\lambda\eta)^2 = 0 \implies \omega_C\lambda^4 - \theta_C\lambda^2 + \frac{\omega_\kappa}{2} = 0\,.
\end{equation}
If we denote $u = \lambda^2$, the characteristic roots are given by
\begin{equation}
    u = \frac{\theta_C \pm \sqrt{\theta_C^2 - 2\omega_C\omega_\kappa}}{2\omega_C}\,, \quad \text{where } \lambda = \pm\sqrt{u}\,.
\end{equation}
Integrating the local phase configuration via $S(\eta) \approx \int \lambda\, \eta\, d\eta = \lambda \eta^2 / 2$, we isolate the fast-mode components $h(\eta) \sim \exp\left( \pm \sqrt{u}\, \eta^2 / 2 \right)$.

\begin{figure}[htbp]
    \centering
    \begin{subfigure}[c]{0.42\textwidth}
        \centering
        \includegraphics[width=\textwidth]{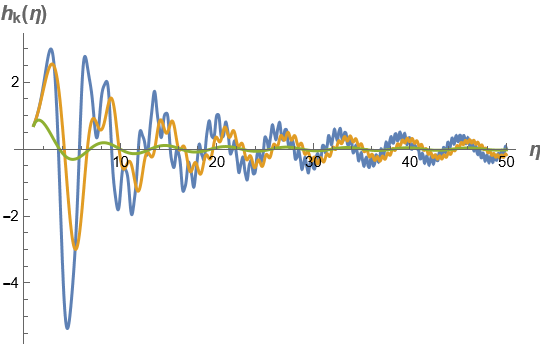}
        \caption{Stable evolution ($k=1$, $\omega_\kappa=2$, $\theta_C=-10$, $\omega_C=10$).}
        \label{fig:Fig5a}
    \end{subfigure}
    \hfill
    \begin{subfigure}[c]{0.42\textwidth}
        \centering
        \includegraphics[width=\textwidth]{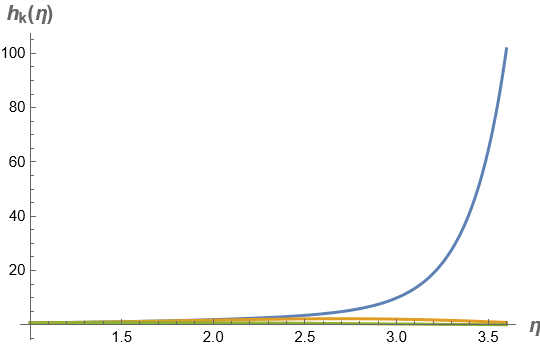}
        \caption{Unstable evolution ($k=1$, $\omega_\kappa=1$, $\theta_C=-2$, $\omega_C=-1$).}
        \label{fig:Fig5b}
    \end{subfigure}
     \hfill
    \begin{subfigure}[t]{0.1\textwidth}
        \includegraphics[width=\textwidth]{Figuras/Legend.png}
    \end{subfigure}
    \caption{Evolution of tensor perturbations in a radiation-dominated background $a(\eta)=\eta$ with initial conditions set at $\eta_i=1$.}
    \label{fig:Fig5}
\end{figure}

Combining both physical branches, the total late-time asymptotic solution for the tensor perturbation evaluates to
\begin{equation}
    h(\eta) \sim \exp\left( \pm \sqrt{ \frac{\theta_C \pm \sqrt{\theta_C^2 - 2\omega_C\omega_\kappa}}{2\omega_C} } \frac{\eta^2}{2} \right) + \frac{1}{\eta}e^{\pm i k\eta}\,.
    \label{eq:solucion_asintotica_6orden}
\end{equation}
Equation~\eqref{eq:solucion_asintotica_6orden} reveals that the field solution diverges exponentially if the parameters allow complex poles, i.e. $\theta_C^2-2\omega_C\omega_\kappa < 0$. In the real pole regime, bounding the solutions to avoid instabilities forces the physical coupling constants to satisfy the parameter space constraints
\begin{equation}
     \begin{cases}
         \theta_C^2-2\omega_C\omega_\kappa > 0 & \text{(Real poles condition)}\,,\\
         \omega_C\,\omega_\kappa > 0\,, & \\
         \theta_C\, \omega_\kappa < 0\,. & 
     \end{cases}   
\end{equation}
Crucially, the final requirement matches the identical condition established for late-time stability under the quadratic gravity subcase. These analytical thresholds show remarkable agreement with the numerical solution displayed in \Cref{fig:Fig5}.

\begin{figure}[htbp]
    \centering
    \begin{subfigure}[c]{0.42\textwidth}
        \centering
        \includegraphics[width=\textwidth]{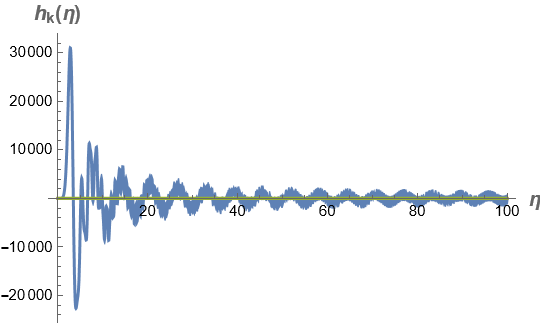}
        \caption{Early initialization ($t_{\textit{i}}=0.1$, $k=1$, $\omega_\kappa=1$, $\theta_C=-2$, $\omega_C=1$).}
        \label{fig:Fig6a}
    \end{subfigure}
    \hfill
    \begin{subfigure}[c]{0.42\textwidth}
        \centering
        \includegraphics[width=\textwidth]{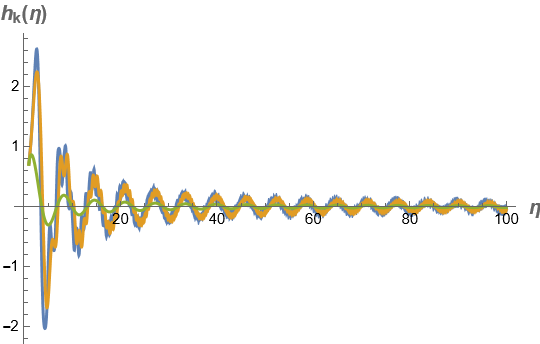}
        \caption{Late initialization ($t_{\textit{i}}=1$, $k=1$, $\omega_\kappa=1$, $\theta_C=-2$, $\omega_C=1$).}
        \label{fig:Fig6b}
    \end{subfigure}
     \hfill
    \begin{subfigure}[t]{0.1\textwidth}
        \includegraphics[width=\textwidth]{Figuras/Legend.png}
    \end{subfigure}
    \caption{Evolution of tensor perturbations in a radiation-dominated background $a(\eta)=\eta$, showcasing the dependence on the initial time scale.}
    \label{fig:Fig6}
\end{figure}

To evaluate the sensitivity of the perturbations to the initial energy scale, we compare the evolution of the modes $h_k(\eta)$ across two distinct initialization regimes, as illustrated in \Cref{fig:Fig6}.

The left plot (\ref{fig:Fig6a}), corresponding to initial conditions set at a very early epoch ($t_{\textit{i}}=0.1$), reveals a discrepancy of up to six orders of magnitude between the sixth-order theory and standard General Relativity. This disparity confirms that the high-energy dynamics is heavily dictated by the higher-derivative operators, which excite the massive ultraviolet degrees of freedom inherent to the theory. 
In contrast, the right plot (\ref{fig:Fig6b}) demonstrates that when the evolution is initiated at a later epoch ($t_{\textit{i}}=1$), the modes amplitudes remain comparable across all different theories. In this scenario, the higher-order corrections lose their relative dominance, acting merely as subleading perturbative corrections to the classical Einstein solution.

\begin{figure}[htbp]
    \centering
    \begin{subfigure}[c]{0.42\textwidth}
        \centering
        \includegraphics[width=\textwidth]{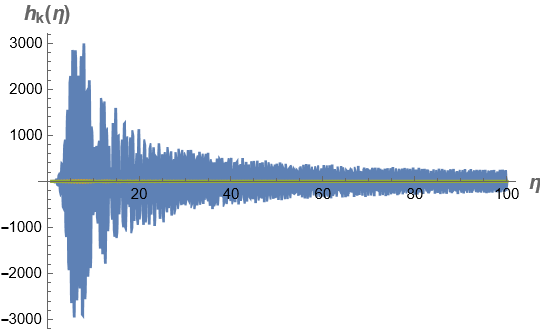}
        \caption{High-energy regime ($k=10$, $\omega_\kappa=1$, $\theta_C=-2$, $\omega_C=1$).}
        \label{fig:Fig7a}
    \end{subfigure}
    \hfill
    \begin{subfigure}[c]{0.42\textwidth}
        \centering
        \includegraphics[width=\textwidth]{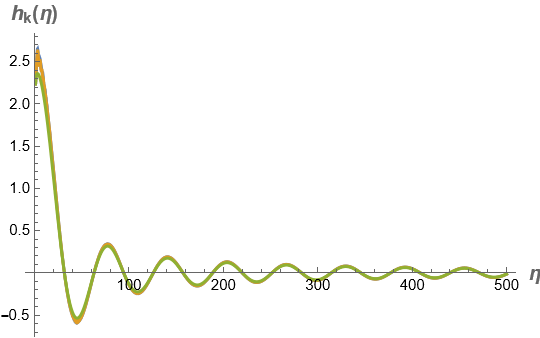}
        \caption{Low-energy regime ($k=0.1$, $\omega_\kappa=1$, $\theta_C=-2$, $\omega_C=1$).}
        \label{fig:Fig7b}
    \end{subfigure}
     \hfill
    \begin{subfigure}[t]{0.1\textwidth}
        \includegraphics[width=\textwidth]{Figuras/Legend.png}
    \end{subfigure}
    \caption{Evolution of tensor perturbations in a radiation-dominated background $a(\eta)=\eta$ across different wavenumbers $k$ with a fixed initialization time $\eta_i=1$.}
    \label{fig:Fig7}
\end{figure}

Although we have shown that the late-time asymptotic stability is independent of the wavenumber $k$, the transient dynamics exhibit a strong sensitivity to this scale. Figure~\ref{fig:Fig7} details this fundamental dependence of the evolution of the perturbation on $k$.

In the high-energy regime shown in the left plot (\ref{fig:Fig7a}), corresponding to $k=10$, the sixth-order theory features oscillations of significantly larger amplitude. This behavior stems from the fact that for large wavenumbers, the higher-derivative terms, which scale as $k^4$ or $k^6$, completely dominate the  equations of motion. 
Conversely, the right plot (\ref{fig:Fig7b}) shows that in the low-energy infrared limit ($k=0.1$), the evolution of the perturbation for the three theories becomes practically indistinguishable. This fact explicitly confirms the smooth decoupling of the higher-derivative degrees of freedom and the precise recovery of General Relativity in the infrared regime.

\subsection{Matter-dominated universe}
For a matter-dominated universe characterized by the scale factor $a(\eta)\propto\eta^2$, this metric does not constitute an exact solution to the background field equations of the sixth-order higher-derivative gravity. However, this cosmological background remains a valid approximate solution in the low-curvature regime, where the linear Einstein-Hilbert term dominates over higher-derivative ones. For Stelle gravity, this approximation holds provided that $|\omega_{\kappa} R| \gg |\theta_{R} R^2|$, which leads to the operational bound
\begin{equation}
    |R| \ll \frac{\omega_{\kappa}}{\theta_{R}}\,.
    \label{eq:condicion_aprox_materia}
\end{equation}
Computing the Ricci scalar for this metric yields $R = 12/\eta^6$. In the deep sub-horizon limit ($k\eta \gg 1$), the curvature scalar decays rapidly, guaranteeing that the low-curvature approximation is robust and self-consistent.

Under this assumption, the full equations of motion for the tensor perturbations in this spacetime evaluate to
\begin{equation}
    \begin{split}
        &\frac{\omega_C}{\eta^4}h^{(VI)}(\eta)-12\frac{\omega_C}{\eta^5}h^{(V)}(\eta)+\left[\frac{\omega_C}{\eta^4}\left( \frac{40}{\eta^2}+3k^2 \right)-\theta_C\right] h^{(IV)}(\eta)+\frac{\omega_C}{\eta^5}\left[ \frac{120}{\eta^2}-24k^2 \right]h'''(\eta) \\
        &\quad +\left[\frac{\omega_C}{\eta^4}\left( -\frac{840}{\eta^4}+80\frac{k^2}{\eta^2}+3k^4 \right)-\frac{\omega_R}{\eta^6}\left( \frac{216}{\eta^2}+84k^2  \right)+\frac{\omega_{\kappa}}{2}\eta^4+\theta_{R}\frac{12}{\eta^2}-2\theta_C k^2 \right]h''(\eta) \\
        &\quad +\left[ -\omega_C\frac{k^2}{\eta^5}\left( \frac{120}{\eta^2}+12k^2 \right)+\frac{\omega_R}{\eta^7}\left( \frac{1728}{\eta^2}+504k^2 \right)+ 2\omega_{\kappa}\eta^3-\theta_{R}\frac{24}{\eta^3} \right] h'(\eta) \\
        &\quad +k^2\left[\frac{\omega_C}{\eta^4}\left( \frac{840}{\eta^4}+40\frac{k^2}{\eta^2}+k^4 \right)+\frac{\omega_R}{\eta^6}\left( 84-\frac{1980}{\eta^2} \right)+\frac{\omega_{\kappa}}{2}\eta^4+\theta_{R}\frac{12}{\eta^2}-\theta_C k^2 \right]h(\eta) \\
        &\quad +\left[ \theta_R\frac{648}{\eta^4}-\omega_R\frac{51840}{\eta^{10}} \right]h(\eta)=0\,.
    \end{split}
\end{equation}
Performing an asymptotic expansion matching the procedure implemented for the radiation era, the dominant equation governing the late-time regime reads
\begin{equation}
    \frac{\omega_C}{\eta^4}h^{(VI)} - 12\frac{\omega_R}{\eta^5}h^{(V)} - \theta_C h^{(IV)} + 24\frac{\omega_C k^2}{\eta^5}h''' + \frac{\omega_{\kappa}}{2}\eta^4 h'' + 2\omega_{\kappa}\eta^3 h' + \frac{\omega_{\kappa}}{2}\eta^4 k^2 h \approx 0\,.
    \label{eq:eom_materia_asintotica}
\end{equation}

Within the quadratic gravity theory, isolating the fourth-order components of ~\eqref{eq:eom_materia_asintotica} leads directly to the analytical late-time solution
\begin{equation}
    h(\eta) \approx C_1 \exp\left( i \sqrt{\frac{-\omega_{\kappa}}{2\theta_C}} \frac{\eta^3}{3} \right) + C_2 \exp\left( -i \sqrt{\frac{-\omega_{\kappa}}{2\theta_C}} \frac{\eta^3}{3} \right) + \frac{1}{\eta^2}\left( C_3 e^{ik\eta} + C_4 e^{-ik\eta} \right)\,.
\end{equation}
Extending this asymptotic balance to the full sixth-derivative theory yields the general solution
\begin{equation}
    h(\eta) \approx C_1 \exp\left[ \pm \sqrt{\frac{1}{2\omega_C}\left( \theta_C \pm \sqrt{\theta_C^2-2\omega_C\omega_\kappa} \right)}\frac{\eta^3}{3} \right] + \frac{1}{\eta^2}\left( C_2 e^{ik\eta} + C_3 e^{-ik\eta} \right)\,.
\end{equation}

\begin{figure}[htbp]
    \centering
    \begin{subfigure}[c]{0.42\textwidth}
        \centering
        \includegraphics[width=\textwidth]{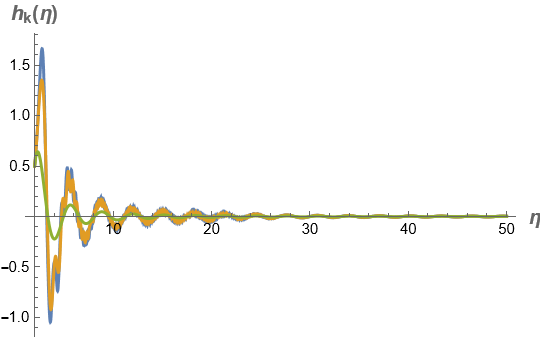}
        \caption{Stable evolution ($k=2$, $\omega_\kappa=1$, $\theta_C=-3$, $\omega_C=2$).}
        \label{fig:Fig8a}
    \end{subfigure}
    \hfill
    \begin{subfigure}[c]{0.42\textwidth}
        \centering
        \includegraphics[width=\textwidth]{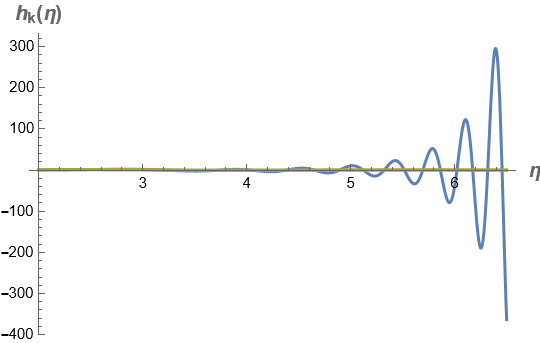}
        \caption{Unstable evolution ($k=2$, $\omega_\kappa=1$, $\theta_C=-3$, $\omega_C=5$).}
        \label{fig:Fig8b}
    \end{subfigure}
     \hfill
    \begin{subfigure}[t]{0.1\textwidth}
        \includegraphics[width=\textwidth]{Figuras/Legend.png}
    \end{subfigure}
    \caption{Evolution of tensor perturbations in a matter-dominated background $a(\eta)=\eta^2$ with a fixed initialization time $\eta_i=1$.}
    \label{fig:Fig8}
\end{figure}

We find out that these fields diverge exponentially under the exact algebraic parameter bounds established during the radiation-dominated era for both gravities. This behavior, illustrated numerically in \Cref{fig:Fig8}, confirms that the late-time stability criterion is uniquely determined by the coupling constants of the fundamental Lagrangian, rather than being an artifact of the specific power-law scaling of a given background expansion of the form $a(t)=t^p$.

\subsection{de Sitter universe}
We now turn to a universe undergoing exponential expansion, characterized by de Sitter space with the conformal scale factor $a(\eta) = -1/(H\eta)$. Provided that the background condition in~\eqref{eq:condicion_hubble_desitter} is satisfied, this metric constitutes an exact solution to the Friedmann equations. Under these conditions, the full equation of motion for the tensor perturbations evaluates to
\begin{equation}
    \begin{split}
        &-\frac{\omega_\Lambda}{6}\eta^2\omega_C h^{(VI)} - \eta\omega_\Lambda\omega_C h^{(V)} + \left[-\frac{\omega_\Lambda}{2}\omega_C\left( k^2\eta^2-\frac{2}{3} \right)-\theta_C \right]h^{(IV)} - 2k^2\eta\omega_\Lambda\omega_C h''' \\
        &\quad + \left[-\frac{k^2}{2}\omega_\Lambda\omega_C\left( k^2\eta^2-\frac{4}{3} \right) + 14k^2\omega_\Lambda\omega_R + \frac{12 \theta_{R}}{\eta^2} - 2 k^2 \theta_C - \frac{3\omega_{\kappa}}{ \eta^2 \omega_\Lambda} \right] h'' \\
        &\quad + \left[ -k^4\eta\omega_\Lambda\omega_C - \frac{24 \theta_{R}}{\eta^3} + \frac{6\omega_{\kappa}}{\eta^3 \omega_\Lambda} \right] h' \\
        &\quad + \left[ -\frac{k^4}{2}\omega_\Lambda\omega_C\left( \frac{k^2}{3}\eta^2-\frac{2}{3} \right) - 14k^2\omega_\Lambda\omega_R + \frac{12 k^2 \theta_{R}}{\eta^2} - k^4 \theta_C - \frac{3 k^2\omega_{\kappa}}{ \eta^2 \omega_\Lambda} \right] h = 0\,.
    \end{split}
    \label{eq:eom_desitter}
\end{equation}

To analyze the asymptotic stability of the perturbations at late cosmological times ($t \to \infty$), we look at the conformal time evolution as $\eta \to 0^-$. In this regime, the physical wavelengths are stretched far beyond the Hubble radius, directly corresponding to the super-horizon limit ($|k\eta| \ll 1$). Extracting the leading-order contributions within each differential coefficient as $\eta \to 0$, the governing equation reduces to
\begin{equation}
    \begin{split}
        &-\frac{\omega_\Lambda\omega_C}{6}\eta^2 h^{(VI)} - \omega_\Lambda\omega_C \eta h^{(V)} + \left(\frac{\omega_\Lambda\omega_C}{3} - \theta_C\right) h^{(IV)} \\
        &\quad + \frac{1}{\eta^2}\left( 12\theta_{R} - \frac{3\omega_{\kappa}}{\omega_\Lambda} \right) h'' - \frac{1}{\eta^3}\left( 24\theta_{R} - \frac{6\omega_{\kappa}}{\omega_\Lambda} \right) h' + \frac{k^2}{\eta^2}\left( 12\theta_{R} - \frac{3\omega_{\kappa}}{\omega_\Lambda} \right) h \approx 0\,.
    \end{split}
    \label{eq:eom_limit_desitter}
\end{equation}
Using the auxiliary constants
\begin{equation}
    \Omega_C = -\frac{\omega_\Lambda\omega_C}{6}, \quad K = -\theta_C - 2\Omega_C, \quad A = 12\theta_{R} - \frac{3\omega_{\kappa}}{\omega_\Lambda}\,,
\end{equation}
we can write the asymptotic super-horizon equation into a compact form
\begin{equation}
    \Omega_C\eta^2 h^{(VI)} + 6\Omega_C\eta h^{(V)} + K h^{(IV)} + \frac{A}{\eta^2} h'' - \frac{2A}{\eta^3} h' + \frac{k^2 A}{\eta^2} h \approx 0\,.
\end{equation}

To extract the scaling behavior of the fields in this limit, we employ a power-law ansatz $h(\eta) \sim \eta^p$. Substituting this into the reduced equation and enforcing a dominant balance analysis as $\eta \to 0$, the spatial gradient term proportional to $k^2 \eta^{p-2}$ is severely suppressed relative to the dominant $\mathcal{O}(\eta^{p-4})$ contributions. Factoring out the leading power of $\eta$, the resulting characteristic indicial polynomial for the exponent $p$ simplifies to
\begin{equation}
    p(p-3) \left[ \Omega_C (p-1)(p-2)(p-4)(p+1) + K(p-1)(p-2) + A \right] = 0\,.
    \label{eq:polinomio_indicial}
\end{equation}

The roots of this polynomial dictate the late-time dynamical fate of the super-horizon modes. 
\subsubsection{Quadratic gravity.}
Restricting our analysis to the fourth-order gravity, the higher-derivative parameter $\Omega_C$ vanishes. The auxiliary coefficients reduce then to
\begin{equation}
    \Omega_C = 0, \quad K = -\theta_C , \quad A = 12\theta_{R} - \frac{3\omega_{\kappa}}{\omega_\Lambda}\,.
\end{equation}
Under these conditions, the characteristic indicial equation yields the exact roots $p_1=0$, $p_2=3$, and $p_{3,4}=\frac{3}{2}\pm \frac{1}{2}\sqrt{\Delta}$, where the discriminant is defined as $\Delta=1+4A/\theta_C$. Consequently, the general asymptotic solution in the super-horizon limit evaluates to
\begin{equation}
    h(\eta) \approx C_1 + C_2 \eta^3 + C_3 \eta^{p_3} + C_4 \eta^{p_4}\,.
\end{equation}

The late-time dynamical stability of these modes is governed by the sign and magnitude of $\Delta$. If $\Delta \leq 0$, the non-trivial roots are either complex or degenerate. In this regime, the real part of the exponent is strictly positive, ensuring that the perturbation either decays to zero or freezes out to a constant amplitude as $\eta \to 0^-$, thereby yielding a stable configuration. In contrast, if $\Delta > 0$, the roots are real and distinct. A divergence in the super-horizon limit occurs if at least one root becomes negative, requiring $\frac{3}{2} < \frac{1}{2}\sqrt{\Delta}$, which mathematically implies that $\Delta > 9$.

\begin{figure}[htbp]
    \centering
    \begin{subfigure}[c]{0.42\textwidth}
        \centering
        \includegraphics[width=\textwidth]{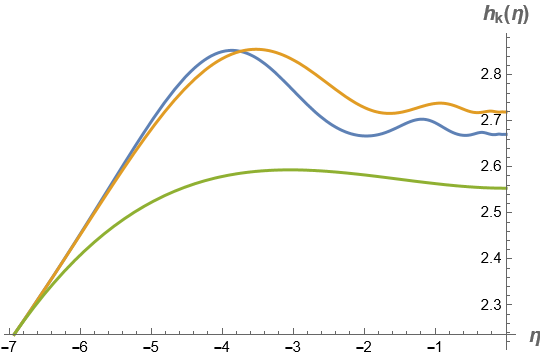}
        \caption{Stable tracking ($k=0.1$, $\omega_\kappa=8$, $\omega_\Lambda=-1$, $\theta_R=-1$, $\theta_C=-9$, $\omega_R=1$, $\omega_C=3$).}
        \label{fig:Fig9a}
    \end{subfigure}
    \hfill
    \begin{subfigure}[c]{0.42\textwidth}
        \centering
        \includegraphics[width=\textwidth]{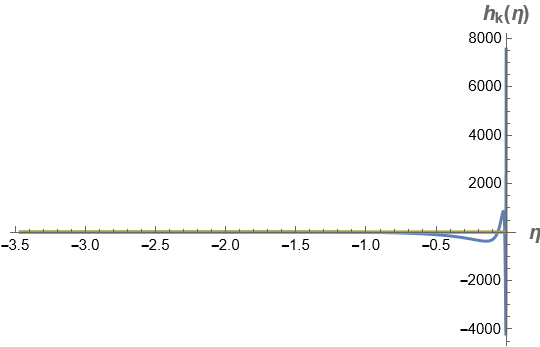}
        \caption{Unstable tracking ($k=2$, $\omega_\kappa=2$, $\omega_\Lambda=-1$, $\theta_R=1/4$, $\theta_C=-1$, $\omega_R=1$, $\omega_C=3$).}
        \label{fig:Fig9b}
    \end{subfigure}
    \hfill
    \begin{subfigure}[t]{0.1\textwidth}
        \includegraphics[width=\textwidth]{Figuras/Legend.png}
    \end{subfigure}
    \caption{Evolution of tensor perturbations in a de Sitter background, illustrating the structural dependence on the theoretical parameter space.}
    \label{fig:Fig9}
\end{figure}

Translating this back to the fundamental coupling constants, the perturbation amplitudes suffer from unstable growth if $1 + 4A/\theta_C > 9$, or equivalently if $A/\theta_C > 2$. Expanding $A$, the analytical instability threshold is explicitly given by
\begin{equation}
    \frac{1}{\theta_C}\left(12\theta_{R}-\frac{3\omega_{\kappa}}{\omega_\Lambda}\right) > 2\,,
\end{equation}
which aligns perfectly with the late-time numerical solution presented in \Cref{fig:Fig9}.
\subsubsection{Sixth-order gravity}
Extending the analysis to the complete sixth-order scenario, we introduce the algebraic substitution $u = p^2 - 3p$. This transforms the bracketed operator in \eqref{eq:polinomio_indicial} into the exact quadratic relation
\begin{equation}
    \Omega_C u^2 - (\theta_C + 4\Omega_C)u + A - 2(\theta_C + 6\Omega_C) = 0\,.
\end{equation}
Solving for the auxiliary variable yields the complex roots
\begin{equation}
    u = \frac{4\Omega_C + \theta_C}{2\Omega_C} \pm \frac{1}{2\Omega_C}\sqrt{\theta_C^2 + 64\Omega_C^2 + 16\theta_C\Omega_C - 4A\Omega_C}\,,
    \label{eq:DefU}
\end{equation}
and inverting to recover the physical exponents $p$ we obtain the four branches as
\begin{equation}
    p = \frac{3}{2} \pm \frac{1}{2}\sqrt{9 + 4u}\,.
\end{equation}
To avoid a divergent amplitude as $\eta \to 0^-$, the physical constraint requires $\text{Re}(p) \geq 0$ for all modes. Therefore, instabilities only manifests if the square root term develops a real component larger than $3$. Specifically, the condition $\text{Re}(\sqrt{9+4u}) > 3$ establishes the instability threshold, which subsequently imposes the necessary constraint $\Im(u)^2 < -9\,\Re(u)$.

\subsection{Singularity-free solutions}
\begin{figure}[H]
    \centering
    \begin{subfigure}[b]{0.49\textwidth}
        \centering
        \includegraphics[width=\textwidth]{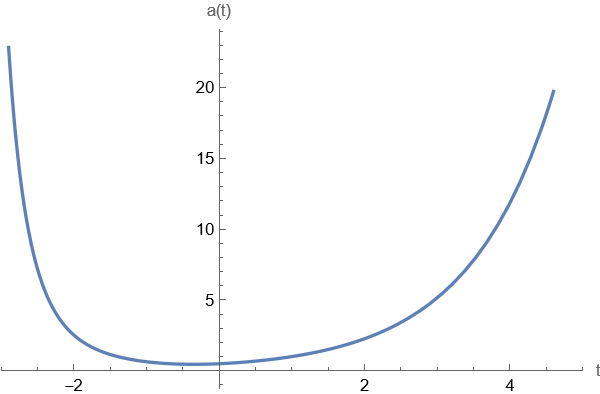}
        \caption{Singularity-free background evolution.}
        \label{fig:Fig11a}
    \end{subfigure}
    \hfill 
    \begin{subfigure}[b]{0.49\textwidth}
        \centering
        \includegraphics[width=\textwidth]{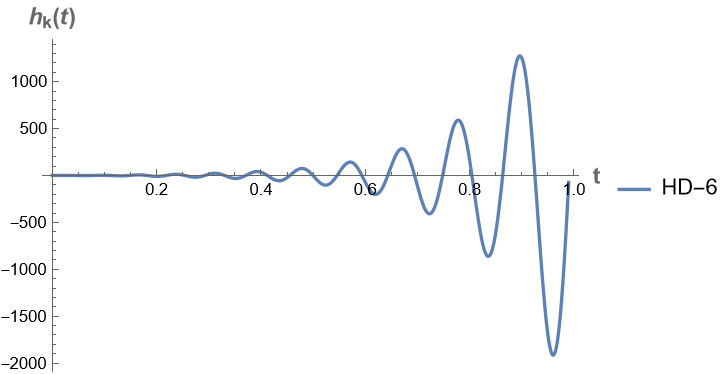}
        \caption{Unstable perturbation dynamics.}
        \label{fig:Fig11b}
    \end{subfigure}
    \caption{Dynamical evaluation of the singularity-free solution and the corresponding temporal evolution of the tensor perturbations within the real pole regime.}
    \label{fig:Fig11}
\end{figure}

\begin{figure}[H]
    \centering
    \begin{subfigure}[b]{0.49\textwidth}
        \centering
        \includegraphics[width=\textwidth]{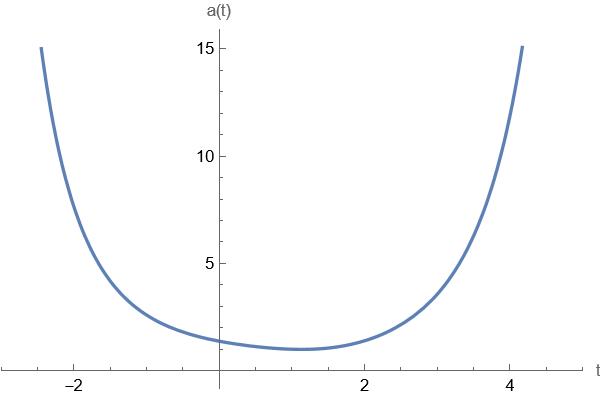}
        \caption{Singularity-free background evolution.}
        \label{fig:Fig12a}
    \end{subfigure}
    \hfill 
    \begin{subfigure}[b]{0.49\textwidth}
        \centering
        \includegraphics[width=\textwidth]{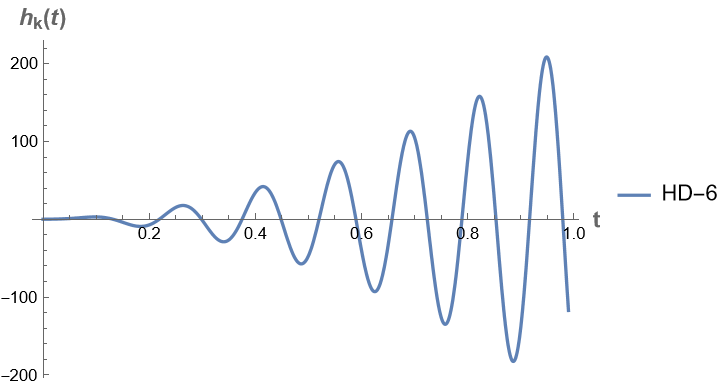}
        \caption{Unstable perturbation dynamics.}
        \label{fig:Fig12b}
    \end{subfigure}
    \caption{Dynamical evaluation of the singularity-free solution and the corresponding temporal evolution of the tensor perturbations within the complex pole regime.}
    \label{fig:Fig12}
\end{figure}

Finally, we investigate the linear stability of the singularity-free cosmological background solutions identified within the complete sixth-order derivative framework (previously illustrated in \Cref{fig:Fig3}). An extensive numerical exploration of the parameter space reveals that, regardless of the specific values chosen for $\theta_C$ and $\omega_C$ (while holding the remaining parameters fixed to guarantee the existence of the background solution), the tensor perturbations experience unbounded growth over time. Consequently, this seems to indicate that these solutions are unstable against tensor-type perturbations, a behavior that is reproduced both in the regime characterized by real poles (\Cref{fig:Fig11}) and in the complex pole scenario (\Cref{fig:Fig12}).

The unstable nature of the found solutions appears to be consistent with their physical origin, since they owe their existence to the excitation of ghost degrees of freedom, which are usually associated with this type of dynamical instability.

\section{Scalar perturbations}
\label{sec:scalar_perturbations}
Let us now apply an analysis  to the evolution of scalar perturbations. We consider a FLRW background metric, linearly perturbed  by a scalar fluctuation parameterized by  the four standard variables ($\phi$, $\psi$, $B$, and $E$), 
\begin{equation}
    ds^2 = -(1+2\phi)dt^2 + 2a(t)\partial_i B \, dt dx^i + a^2(t)\left[(1-2\psi)\delta_{ij} + 2\partial_i\partial_j E \right]dx^idx^j \,.
\end{equation}

To derive the linearized equations of motion for these scalar perturbations, we expand the total action to second order in the metric fluctuations. By applying the Euler-Lagrange formalism we get a coupling of the second order perturbed geometric sector  to the first-order fluctuations of the energy-momentum tensor
\begin{equation}
    \begin{split}
        \delta^{(2)}S &= \int d^4x \left[ -\delta\left( \sqrt{-g}E^{\mu\nu}\right) + \delta\left( \sqrt{-g}T^{\mu\nu} \right) \right]\delta g_{\mu \nu} \\
        &= \int d^4x \left[ -\delta\left( \sqrt{-g}E^{00}\right) + \delta(\sqrt{-g})T^{00} + \sqrt{-g}\delta T^{00} \right](-2\phi) \\
        &\quad + \int d^4x \left[ -\delta\left( \sqrt{-g}E^{0i}\right) + \delta(\sqrt{-g})T^{0i} + \sqrt{-g}\delta T^{0i} \right](2a\partial_i B) \\
        &\quad + \int d^4x \left[ -\delta\left( \sqrt{-g}E^{ij}\right) + \delta(\sqrt{-g})T^{ij} + \sqrt{-g}\delta T^{ij} \right](-2a^2)\left( \psi\delta_{ij} - \partial_i\partial_j E \right) \,.
    \end{split}
\end{equation}

To isolate the physical dynamics and eliminate residual coordinate artifacts, we adopt the Newtonian (or longitudinal) gauge. This choice naturally suppresses the scalar shear and off-diagonal spatial components by enforcing $B=0$ and $E=0$. Consequently, the scalar sector of the metric is completely characterized by two physical variables: the Newtonian potential $\phi$ and the spatial curvature perturbation $\psi$.

Computing  the variations with respect to the metric variables and imposing these gauge constraints we get the complete set of linearized field equations, where the variations with respect to the spatial shifts provide the corresponding momentum and anisotropic stress constraints
\begin{equation}
    \begin{aligned}
        \mathcal{E}_\phi &= 2\delta\left( \sqrt{-g}E^{00}\right) + 2\rho(t)a^3(t)(\phi+3\psi) = 0 \,,\\
        \mathcal{E}_B &= 2\delta\left( \sqrt{-g}E^{0i}\right) = 0 \,,\\
        \mathcal{E}_\psi &= 2a^2(t)\delta_{ij}\delta\left( \sqrt{-g}E^{ij}\right) + 6P(t)a^4(t)(\psi-\phi) = 0 \,,\\
        \mathcal{E}_E &= -2a^2(t)\partial_i\partial_j\delta\left( \sqrt{-g}E^{ij}\right) + 2P(t)a^3(t)k^2(\psi-\phi) = 0 \,.
    \end{aligned}
\end{equation}

\subsection{Minkowski background}
In a flat Minkowski background, the linearized equations of motion for the scalar perturbations simplify significantly. The system reduces to two independent differential equations governing the evolution of the Newtonian potential $\phi$ and the spatial curvature perturbation $\psi$, given by
\begin{align}
    & -4 k^2 \omega_\kappa \psi + \theta_C \left(-\frac{8}{3} k^4 \phi + \frac{8}{3} k^4 \psi\right) - \frac{8}{3} k^4 \omega_C \left(k^2 \phi + k^2 \psi + \phi'' + \psi''\right) \nonumber \\
    & \quad + \theta_R \left(8 k^4 \phi - 16 k^4 \psi - 24 k^2 \psi''\right) + 4 \omega_R \left(-2 k^6 \phi + 4 k^6 \psi + 2 \left(-k^4 \phi'' + 5 k^4 \psi'' + 3 k^2 \psi^{(4)}\right)\right)=0 \,, \\
    \nonumber \\
    & \theta_C \left(-\frac{8}{3} k^4 \phi + \frac{8}{3} k^4 \psi\right) - \frac{8}{3} k^4 \omega_C \left(k^2 \phi + k^2 \psi + \phi'' + \psi''\right) + \omega_\kappa \left(-4 k^2 \phi + 4 k^2 \psi + 12 \psi''\right) \nonumber \\
    & \quad + \theta_R \left(-16 k^4 \phi + 32 k^4 \psi - 24 k^2 \phi'' + 96 k^2 \psi'' + 72 \psi^{(4)}\right) \nonumber \\
    & \quad - 4 \omega_R \left(-4 k^6 \phi + 8 k^6 \psi + 2 \left(-5 k^4 \phi'' + 16 k^4 \psi'' + 3 \left(-k^2 \phi^{(4)} - 7 k^2 \psi^{(4)}\right) + 3 \psi^{(6)}\right)\right)=0 \,.
\end{align}

\subsubsection{Quadratic gravity}
In quadratic gravity, the higher-order couplings $\omega_C$ and $\omega_R$ vanish. The system can be decoupled into a single, fourth-order differential equation governing the curvature perturbation $\psi$
\begin{equation}
    \psi^{(IV)}+\left( k^2+\frac{\omega_\kappa}{6\theta_R}-\frac{\omega_\kappa}{2\theta_C} \right)\psi''+\left( k^4+k^2\left( \frac{\omega_\kappa}{6\theta_R}-\frac{\omega_\kappa}{2\theta_C} \right)-\frac{\omega_\kappa^2}{12\theta_R\theta_C} \right)\psi=0 \,,
\end{equation}
where the primes denote derivatives with respect to the cosmological time $t$. This evolution equation factors into a product of two distinct wave operators, exhibiting the physical propagation of the massive scalar and massive spin-2 degrees of freedom,
\begin{equation}
    \left( \Box -m_1^2 \right)\left( \Box -m_2^2 \right)\psi=0\,,
\end{equation}
where the masses are defined as
\begin{equation}
    m_1^2=\frac{\omega_\kappa}{6\theta_R}\,,\quad m_2^2=-\frac{\omega_\kappa}{2\theta_C} \,.
\end{equation}

Integrating this decoupled system yields the general analytical solution for $\psi(t)$. Substituting this result back into the fundamental field equations allows us to solve  the Newtonian potential $\phi(t)$ equation of motion, yielding the complete solution
\begin{align}
    \psi(t) &= c_1 e^{-it \sqrt{k^2+m_1^2}} + c_2 e^{it \sqrt{k^2+m_1^2}} + c_3 e^{-it \sqrt{k^2+m_2^2}} + c_4 e^{it \sqrt{k^2+m_2^2}} \,, \\
    \phi(t) &= \frac{1}{2} \left[ -2 c_1 e^{-it \sqrt{k^2+m_1^2}} - 2 c_2 e^{it \sqrt{k^2+m_1^2}} + c_3 \left(-2 + \frac{3 \omega_\kappa}{k^2 \theta_C}\right)  e^{-it \sqrt{k^2+m_2^2}} + c_4 \left(-2 + \frac{3 \omega_\kappa}{k^2 \theta_C}\right)  e^{it \sqrt{k^2+m_2^2}}  \right] \,.
\end{align}

To ensure physical stability and avoid tachyonic or exponential growth within the scalar sector, the propagating frequencies must remain strictly real, which leads to the following stability bounds on the comoving wavenumber $k$,
\begin{equation}
    k^2 > -\frac{\omega_\kappa}{6\theta_R}\,,\quad k^2 > \frac{\omega_\kappa}{2\theta_C} \,.
\end{equation}
\subsubsection{Sixth-order gravity}
For the sixth-order super-renormalizable gravity theory, the decoupled equation governing the evolution of the curvature perturbation $\psi(t)$ factors into a product of four distinct wave operators satisfying the equation
\begin{equation}
    \left(\Box-m_{(0)+}^2 \right) \left(\Box-m_{(0)-}^2 \right) \left(\Box-m_{(2)+}^2 \right) \left(\Box-m_{(2)-}^2 \right)\psi = 0\,,
\end{equation}
where $m_{(0)\pm}$ and $m_{(2)\pm}$ denote the effective mass scales of the physical spin-0 (\Cref{eq:MasasEscalares}) and spin-2 poles (\Cref{eq:tensormass6Hd}), respectively. The general analytical solution to this higher-derivative differential system is expressed as a linear combination of plane waves corresponding to the excitations of each mass sector
\begin{equation}
    \psi(t) = \sum_{\lambda=\pm} \left( C_{1,\lambda} e^{-it\sqrt{k^2+m_{(0)\lambda}^2}} + C_{2,\lambda} e^{it\sqrt{k^2+m_{(0)\lambda}^2}} + C_{3,\lambda} e^{-it\sqrt{k^2+m_{(2)\lambda}^2}} + C_{4,\lambda} e^{it\sqrt{k^2+m_{(2)\lambda}^2}} \right)\,.
\end{equation}
An equivalent superposition of comoving modes also governs the temporal evolution of the Newtonian potential $\phi(t)$. 

To guarantee physical stability and avoid tachyonic exponential divergences within the scalar sector, the propagating frequencies must remain strictly real. This requires the comoving wavenumber to satisfy $k^2 > -m_{(2)\pm}^2$ and $k^2 > -m_{(0)\pm}^2$. In terms of the fundamental coupling parameters of the Lagrangian, these stability thresholds are given by
\begin{align}
    k^2 &> -\frac{\theta_C \mp \sqrt{\theta_C^2 - 2\omega_C\omega_\kappa}}{2\omega_C} \,, \label{eq:spin2_bound} \\
    k^2 &> \frac{3\theta_R \mp \sqrt{9\theta_R^2 + 6\omega_R\omega_\kappa}}{6\omega_R} \,, \label{eq:spin0_bound}
\end{align}
where the right-hand sides correspond to the boundaries of the spin-2 and spin-0 sectors, respectively.

It is important to note that if the parameters of the Lagrangian are chosen such that the theory presents complex poles, the solutions will inevitably diverge. Specifically, if the arguments of the inner square roots are negative, the effective masses squared become complex quantities. As a consequence, the corresponding physical frequencies will acquire a non-zero imaginary component. This introduces a real exponential factor in the temporal evolution of the mode amplitudes, leading to uncontrolled exponential growth. Therefore, in the complex pole regime, the Minkowski background becomes unstable under scalar perturbations, regardless of the value of the wavenumber \( k \).

\subsection{de Sitter background}
Let us now analyze the evolution of scalar perturbations within a de Sitter spacetime with  scale factor $a(t) \propto e^{Ht}$. Following the approach of Reference \cite{DeFelice:2023psw}, by algebraically manipulating the set of four coupled equations governing the dynamics of the fluctuations we can reduce them to independent higher-order ordinary differential equations for a single variable, which can be chosen as either the Newtonian potential $\phi$ or the spatial curvature perturbation $\psi$.

\subsubsection{Quadratic gravity}
In quadratic gravity, the decoupled evolution equations for the scalar potentials reduce to fourth-order ordinary differential equations of the form
\begin{align}
    \psi^{(IV)} + \mu_1\dddot{\psi} + \mu_2\ddot{\psi} + \mu_3\dot{\psi} + \mu_4\psi &= 0\,, \\
    \phi^{(IV)} + \nu_1\dddot{\phi} + \nu_2\ddot{\phi} + \nu_3\dot{\phi} + \nu_4\phi &= 0\,,
\end{align}
where the coefficients $\mu_i$ and $\nu_i$ are generally complex, time-dependent functions of the background kinematics. The stability of the solution can easily be obtained from  of the analysis of  these relations in the asymptotic sub-horizon and super-horizon regimes.

\paragraph{Sub-Horizon limit ($k/(aH) \gg 1$)}
In this regime, the physical wavelength of a given perturbation mode is heavily suppressed with respect  to the Hubble radius. Therefore, the leading-order behavior of the time-dependent coefficients takes the form
\begin{equation}
    \begin{aligned}
        \mu_1 = \nu_1 \approx 6H\,,\quad \mu_2 = \nu_2 \approx 2e^{-2Ht}k^2\,,\quad \mu_3 = \nu_3 \approx 2e^{-2Ht}Hk^2\,,
        \quad\mu_4 = \nu_4 \approx e^{-4Ht}k^4\,, 
    \end{aligned}
\end{equation}
which allows a factorization of the fourth-order operator, reducing the system to $\Box^2\psi \approx 0$. 

By explicitly expanding the covariant operator $\Box\psi = \ddot{\psi} + 3H\dot{\psi} + (k^2/a^2)\psi$ and changing to the dimensionless conformal time coordinate $x \equiv k/(aH) = (k/H)e^{-Ht}$, the equation reads
\begin{equation}
    \Box^2\psi = H^4\left( x^2\frac{d^2\psi}{dx^2} - 2x\frac{d\psi}{dx} + x^2\psi \right)^2 \equiv H^4 \mathcal{L}^2\psi\, \approx0,
\end{equation}
where $\mathcal{L} = x^2 \frac{d^2}{dx^2} - 2x\frac{d}{dx} + x^2$. To obtain the scaling behavior when $x \gg 1$, we employ a WKB ansatz $\psi(x) = x^p e^{\pm ix}$. By retaining only the leading-order terms in the high-frequency limit, the iterated differential operator acts as
\begin{equation}
    \mathcal{L}^2(x^p e^{\pm ix}) = (\pm 2i)^2 p(p-1) \, x^{p+2} e^{\pm ix} \approx 0\,,
\end{equation}
whose roots are $p = 0$ and $p = 1$. Mapping these modes back to physical cosmological time, the general asymptotic sub-horizon solution evaluates to
\begin{equation}
    \psi(t) = \left[ C_{1,2} + C_{3,4}\left(\frac{k}{H}e^{-Ht}\right) \right] e^{\pm i \frac{k}{H}e^{-Ht}}\,,
\end{equation}
which represent oscillatory modes that remain bounded inside the horizon.

\paragraph{Super-Horizon limit ($k/(aH) \ll 1$)}
When physical wavelengths are stretched beyond the Hubble horizon, the terms proportional to the comoving wavenumber $k$ are highly suppressed. In this super-horizon regime, the time-dependent coefficients freeze out into the constant parameter values
\begin{align}
    \mu_1 &= \nu_1 \approx 2H\,, \\
    \mu_2 &= \nu_2 \approx \frac{-18H^2\theta_R(\theta_C+4\theta_R)+(\theta_C-3\theta_R)\omega_\kappa}{6\theta_C\theta_R}\,, \\
    \mu_3 &= \nu_3 \approx \frac{-H\left[\theta_C\omega_\kappa+9\theta_R(24H^2\theta_R+\omega_\kappa)\right]}{6\theta_C\theta_R}\,, \\
    \mu_4 &= \nu_4 \approx -\frac{\omega_\kappa(24H^2\theta_R+\omega_\kappa)}{12\theta_C\theta_R}\,.
\end{align}
Since these coefficients are stationary, the system admits an exact exponential solution of the form $\psi(t) \sim e^{\lambda t}$. Solving the corresponding fourth-order characteristic polynomial yields to the general solution
\begin{equation}
    \psi(t) = C_1 e^{\frac{H}{2}(1-\Omega)t} + C_2 e^{\frac{H}{2}(1+\Omega)t} + C_3 e^{-\frac{3H}{2}(1-\Omega)t} + C_4 e^{-\frac{3H}{2}(1+\Omega)t}\,,
\end{equation}
where the dimensionless auxiliary factor $\Omega$ is defined as
\begin{equation}
    \Omega = \sqrt{\frac{\theta_C+48\theta_R}{\theta_C} + \frac{2\omega_\kappa}{H^2\theta_C}}\,.
\end{equation}
An analysis of these characteristic exponents reveals that, independently of the specific choice of the fundamental gravitational couplings ($\theta_C, \theta_R, \omega_\kappa$), the real part of at least one eigenvalue $\lambda$ is strictly positive. Consequently, after crossing outside the Hubble horizon, scalar perturbation modes inevitably undergo unbounded exponential growth. This shows that the de Sitter vacuum within the standard quadratic gravity formulation is unstable under  scalar fluctuations.

\subsubsection{Sixth-order gravity}
In sixth-order gravity, the decoupled evolution equations of the scalar fluctuations give rise to the following eighth-order ordinary differential system. 
%This is associated to the dynamics of the four independent scalar degrees of freedom propagating within the theory, satisfying
\begin{align}
    \psi^{(VIII)} + \mu_1\psi^{(VII)} + \mu_2\psi^{(VI)} + \mu_3\psi^{(V)} + \mu_4\psi^{(IV)} + \mu_5\dddot{\psi} + \mu_6\ddot{\psi} + \mu_7\dot{\psi} + \mu_8\psi &= 0\,, \\
    \phi^{(VIII)} + \nu_1\phi^{(VII)} + \nu_2\phi^{(VI)} + \nu_3\phi^{(V)} + \nu_4\phi^{(IV)} + \nu_5\dddot{\phi} + \nu_6\ddot{\phi} + \nu_7\dot{\phi} + \nu_8\phi &= 0\,.
\end{align}

\paragraph{Sub-Horizon limit ($k/(aH) \gg 1$)}
In the sub-horizon regime, where physical momentum dominates over the background curvature, the leading-order contributions to the time-dependent coefficients reduce to 
\begin{align}
    \mu_1 &= \nu_1 \approx 12H\,, & \mu_2 &= \nu_2 \approx 4e^{-2Ht}k^2\,, & \mu_3 &= \nu_3 \approx 12e^{-2Ht}Hk^2\,, \nonumber \\
    \mu_4 &= \nu_4 \approx 6e^{-4Ht}k^4\,, & \mu_5 &= \nu_5 \approx -12e^{-4Ht}Hk^4\,, & \mu_6 &= \nu_6 \approx 4e^{-6Ht}k^6\,, \nonumber \\
    \mu_7 &= \nu_7 \approx -12e^{-6Ht}Hk^6\,, & \mu_8 &= \nu_8 \approx e^{-8Ht}k^8\,.
\end{align}
This algebraic structure implies that the eighth-order differential operator  again  factorizes  asymptotically into d'Alembertian operators acting on the fields, $\Box^4\phi \approx 0$. Mapping this system to the dimensionless conformal momentum coordinate $x$ via the differential operator $\mathcal{L}$ introduced in the previous quadratic gravity analysis, the sub-horizon evolution equation becomes $H^8\mathcal{L}^4\phi \approx 0$.

Introducing the high-frequency WKB ansatz $\phi(x) = x^p e^{\pm ix}$ and restricting the   analysis to the asymptotic regime  $x \gg 1$, the characteristic polynomial expression leads  to
\begin{equation}
    \mathcal{L}^4\left(x^p e^{\pm ix}\right) \approx (\pm 2i)^4 (p-1)p(p+1)(p+2)x^{p+4}e^{\pm ix} = 0\,.
\end{equation}
From the  leading terms we get imposes constraints on the exponent, yielding the roots $p \in \{-2, -1, 0, 1\}$. Reverting to physical cosmological time, the general sub-horizon solution is spanned by the eight independent modes
\begin{equation}
    \phi(t) = \left[ C_{1,2} + C_{3,4} \left(\frac{k}{H}e^{-Ht}\right) + C_{5,6} \left(\frac{H}{k}e^{Ht}\right) + C_{7,8} \left(\frac{H^2}{k^2}e^{2Ht}\right) \right] e^{\pm i \frac{k}{H}e^{-Ht}}\,.
\end{equation}
While the branches for $p=0$ and $p=1$ remain bounded, the modes corresponding to negative powers ($p=-1$ and $p=-2$) exhibit rapid exponential growth inside the horizon. This behavior reveals a fundamental departure from quadratic gravity, where sub-horizon scalar modes remain oscillatory stable. The inclusion of sixth-order derivative terms introduces an ultraviolet instability well before  horizon crossing.

\paragraph{Super-Horizon limit ($k/(aH) \ll 1$)}
As modes expand far beyond the Hubble radius, the spatial gradients become subdominant and the time-dependent coefficients freeze out at the constant parameter values
\begin{align}
    \mu_1 &= \nu_1 \approx 4H\,, \nonumber \\
    \mu_2 &= \nu_2 \approx -14H^2 - \frac{\theta_C}{\omega_C} - \frac{\theta_R}{\omega_R}\,, \nonumber \\
    \mu_3 &= \nu_3 \approx -56H^3 - \frac{5H\theta_C}{\omega_C} - \frac{H\theta_R}{\omega_R}\,, \nonumber \\
    \mu_4 &= \nu_4 \approx \frac{222 H^4 \omega_C \omega_R + H^2 \theta_R (78 \omega_C + 72 \omega_R) + 6 \theta_C (\theta_R + H^2 \omega_R) - (\omega_C - 3 \omega_R) \omega_\kappa}{6 \omega_C \omega_R}\,, \nonumber \\
    \mu_5 &= \nu_5 \approx \frac{H \left[ 372 H^4 \omega_C \omega_R + 3 H^2 \theta_R (13 \omega_C + 72 \omega_R) + \theta_C (6 \theta_R + 51 H^2 \omega_R) + (\omega_C + 9 \omega_R) \omega_\kappa \right]}{3 \omega_C \omega_R}\,, \nonumber \\
    \mu_6 &= \nu_6 \approx \frac{-576 H^6 \omega_C \omega_R - 72 H^4 (2 \theta_R \omega_C + \theta_C \omega_R - 5 \theta_R \omega_R) + (\theta_C - 3 \theta_R) \omega_\kappa + H^2 \Gamma}{6 \omega_C \omega_R}\,, \nonumber \\
    \mu_7 &= \nu_7 \approx -\frac{H \left[ 864 H^4 \theta_R \omega_R + (\theta_C + 9 \theta_R) \omega_\kappa + 4 H^2 (54 \theta_R^2 + 2 \omega_C \omega_\kappa + 9 \omega_R \omega_\kappa) \right]}{6 \omega_C \omega_R}\,, \nonumber \\
    \mu_8 &= \nu_8 \approx -\frac{\omega_\kappa (24 H^2 \theta_R - \omega_\kappa)}{12 \omega_C \omega_R}\,,
\end{align}
where we have defined $\Gamma \equiv -18 \theta_C \theta_R - 72 \theta_R^2 + 7 \omega_C \omega_\kappa + 15 \omega_R \omega_\kappa$. In this stationary limit, the corresponding eighth-order characteristic polynomial decouples into two distinct eigenvalue sectors 
\begin{equation}
    \phi(t) = \sum_{j=1}^4 C_j e^{\lambda_{R,j} t} + \sum_{j=5}^8 C_j e^{\lambda_{C,j} t}\,,
\end{equation}
where the characteristic roots are parameterized by 
\begin{align}
    \lambda_{R} &= -\frac{3H}{2} \pm \frac{\sqrt{3}}{6} \sqrt{ \frac{6\theta_R + 51H^2\omega_R \pm \sqrt{36(\theta_R + 4H^2\omega_R)^2 + 24\omega_R\omega_\kappa}}{\omega_R} }\,, \\
    \lambda_{C} &= \frac{H}{2} \pm \frac{1}{2} \sqrt{ \frac{2\theta_C + 17H^2\omega_C \pm \sqrt{4(\theta_C + 8H^2\omega_C)^2 - 8\omega_C(24H^2\theta_R - \omega_\kappa)}}{\omega_C} }\,.
\end{align}
A simple evaluation of these roots confirms that an infrared super-horizon instability remains unavoidable. Specifically, the real part of at least one resulting eigenvalue remains strictly positive and, consequently, scalar perturbation modes cross  the Hubble horizon with an unbounded exponential growth, showing that the de Sitter background is unstable for all scales within the sixth-order gravity theory.

\begin{figure}[H]
    \centering
    \includegraphics[width=0.75\textwidth]{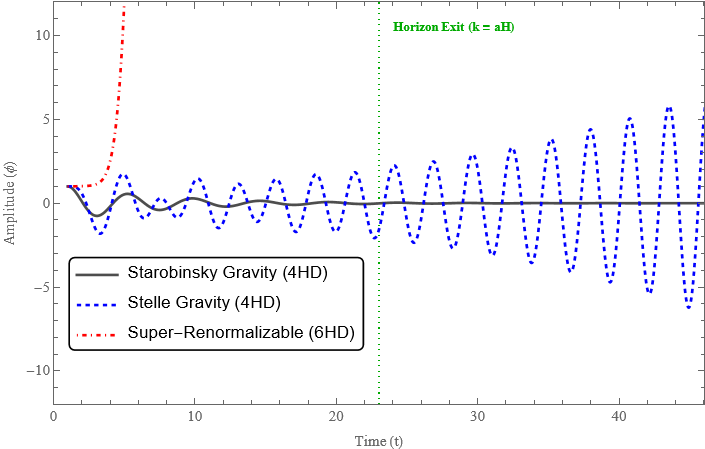}
    \caption{Evolution of the scalar perturbation $\phi(t)$ in a de Sitter background. The vertical dash-dotted line denotes the horizon ($k = aH$). The Starobinsky model reaches a stable freeze-in, while Stelle gravity suffers from super-horizon instabilities. The super-renormalizable sixth-derivative theory (6HD) exhibits a global divergence that grows exponentially within the horizon.}
    \label{fig:PertEscalaresdSitter}
\end{figure}

These semi-analytical results closely match our numerical simulations, as illustrated in \Cref{fig:PertEscalaresdSitter}, which shows the time evolution of the scalar perturbations in a de Sitter background  for  different higher derivatives gravitational theories. For the Starobinsky case, the perturbation oscillates initially and subsequently stabilizes. In contrast, in quadratic gravity, the mode remains bounded at the sub-horizon regime but diverges rapidly once it crosses the horizon. Finally, for sixth-order derivative theory, the perturbation exhibits a severe instability, growing exponentially well before horizon crossing.

\iffalse
Additionally, we investigated the numerical behavior of scalar modes during the radiation-dominated era ($a(t) \propto t^{1/2}$). In this regime, the explicit coupling between the higher-derivative gravitational sectors and the relativistic fluid energy-momentum tensor drastically alters the stability landscape. Our numerical solutions indicate that for both the Stelle and sixth-order derivative frameworks, the scalar perturbations experience an immediate, severe divergence. Rather than showcasing standard acoustic oscillations, the higher-derivative degrees of freedom trigger rapidly growing ghost-like instabilities in the presence of radiation, rendering the perturbation theory unstable in this epoch.
\fi

\section{Effective Field Theory (EFT) approach to higher-derivative cosmological perturbations}
\label{sec:EFT_observables}

As shown in  previous sections,   higher-derivative theories  leads to ghost-like instabilities in the scalar perturbation sector. These instabilities prevent the standard quantization procedure deep inside the horizon, specifically, the establishment of the Bunch-Davies vacuum and thereby prevent a reliable calculation of the primordial scalar and tensor power spectra predicted by inflation.

To overcome this pathology and extract meaningful physical observables, we adopt a different approach based on the EFT framework. By treating the benign Starobinsky $R^2$ term as fundamental in the action and embedding the remaining higher-derivative terms as suppressed perturbative corrections, the theory now circumvents the introduction of spurious Ostrogradsky ghosts \cite{RomeroCastellanos:2018inv,Bianchi:2025tyl}. This prescription ensures that the well-behaved scalaron degree of freedom inherent to $f(R)$ gravity is preserved, allowing the use of standard cosmological perturbation theory to evaluate how these higher-order UV corrections shift the inflationary observables.

The implementation of the EFT scheme requires an order-by-order reduction at the level of the perturbed action. Initially, the second-order variation of the total action is computed with respect to both the scalar and tensor metric fluctuations. Generically, the higher-derivative operators introduce time derivatives of second and higher orders into the perturbation equations of motion. To systematically eliminate these higher order modes, we perturbatively substitute the lower-order background equations of motion derived exclusively from the dominant Starobinsky action. Through this iterative removal of higher-order time derivatives, the second-order action is mapped onto a canonical form depending only on the perturbation fields and their first derivatives. 

Consequently, the physical modes can be canonically quantized deep inside the sub-horizon regime. The power spectra is then evaluated at the horizon crossing ($k = aH$), where the frozen-out mode amplitudes can be related with inflationary parameters such as the scalar spectral index $n_s$ and the tensor-to-scalar ratio $r$. 

A particular feature of this EFT framework concerns the six-derivative scalar operator, $R\Box R$. By taking the trace of the field equations generated by the basic Starobinsky action, we get at leading-order dynamical equation for the Ricci scalar is verified,
\begin{equation}
    \Box R = \frac{\omega_\kappa}{6\theta_R} R\,.
\end{equation}
By substituting this on-shell relation directly back into the higher-derivative sector of the action \eqref{eq:S6}, the relevant terms reduce to
\begin{equation}
    \theta_R R^2 + \omega_R R\Box R = \left( \theta_R + \frac{\omega_R\omega_\kappa}{6\theta_R} \right) R^2\,.
\end{equation}
Therefore, within the validity range of the EFT, the inclusion of the operator $R\Box R$ is completely equivalent to a constant shift of the quadratic Ricci coupling $\theta_R$. Since this does not generate additional degrees of freedom because it simply shifts the effective mass scale of the Starobinsky scalaron, its effects can be absorbed into the original action parameters without loss of generality. For the rest of this perturbative analysis, the $R\Box R$ term will be accordingly omitted.
\subsubsection{Scalar power spectrum and canonical quantization}
\label{subsect:ScalarPowerSpectrum}

Let us start the new analysis of the scalar sector by isolating the Starobinsky contribution to the total action. The reduction procedure is formulated as follows. First, the second-order variation of the action is performed within the conformal Newtonian gauge. Using standard techniques, such as the introduction of auxiliary fields acting as Lagrange multipliers and the subsequent integration over non-dynamical degrees of freedom, the action can be written in terms of the gauge-invariant comoving curvature perturbation $\mathcal{R}$, defined as
\begin{equation}
    \mathcal{R} = \psi + \frac{\mathcal{H}}{\bar{R}'}\delta R\,,
\end{equation}
where $\mathcal{H} \equiv a'/a$ denotes the conformal Hubble parameter and $\delta R$ represents the linear perturbation of the Ricci scalar. In the Newtonian gauge, the latter takes the explicit form
\begin{equation}
    \delta R = \frac{1}{a^2(\eta)} \left[ -6\psi''(\eta) - 6\mathcal{H}\phi'(\eta) - 18\mathcal{H}\psi'(\eta) - 12(\mathcal{H}' + \mathcal{H}^2)\phi(\eta) - 4k^2\psi(\eta) + 2k^2\phi(\eta) \right]\,.
\end{equation}
By substituting this trace fluctuation back into the quadratic action, the Starobinsky sector reads
\begin{equation}
    S_{\text{Starobinsky}} = \int d\eta \, \frac{12 \left( -\frac{1}{2} k^2 \mathcal{R}^2 + \frac{1}{2} \mathcal{R}'^2 \right) \left( \omega_\kappa a^3 + 12 \theta_R a'' \right) \left( \omega_\kappa a^3 a'^2 + 12 \theta_R a'^2 a'' - 6 \theta_R a a''^2 \right)^2}{a \left( \omega_\kappa a^3 a'^2 + 12 \theta_R a'^2 a'' + 6 \theta_R a a''^2 \right)^2}\,.
    \label{eq:ActionScalarStarobinsky}
\end{equation}
Next, using the background field equations, we express the second-order variation of the remaining higher-derivative terms as functions of $\mathcal{R}$. These operators introduce higher-order time derivatives, namely $\mathcal{R}''$ and $\mathcal{R}'''$. To eliminate them, we perform a perturbative reduction of order by substituting the lower-order equations of motion derived from the action \eqref{eq:ActionScalarStarobinsky}. This procedure yields an effective quadratic action depending solely on $\mathcal{R}$ and its first time derivative $\mathcal{R}'$. Introducing the canonical Mukhanov-Sasaki variable $v = z\mathcal{R}$, the effective action can be written into its canonical form
\begin{equation}
    S_{\text{HD}} = \int d\eta \, \frac{1}{2} \left[ v'^2 - c_s^2(\eta) k^2 v^2 + \frac{z''}{z} v^2 \right]\,,
\end{equation}
where the time-dependent effective speed of sound $c_s(\eta)$ is evaluated to be
\begin{equation}
\begin{split}
   c_s^2(\eta) =\;& 1 - \frac{2 \theta_C \left( \omega_\kappa a^3 a'^2 + 12 \theta_R a'^2 a'' - 6 \theta_R a a''^2 \right)^2}{a a'^2 \left( \omega_\kappa a^3 + 12 \theta_R a'' \right)^3} \\
   & + \frac{\omega_C \mathcal{N}}{3 \theta_R a'^4 \left( \omega_\kappa a^4 + 12 \theta_R a a'' \right)^5 \left( \omega_\kappa a^3 a'^2 + 12 \theta_R a'^2 a'' + 6 \theta_R a a''^2 \right)^2}\,,
\end{split}
\end{equation}
where the polynomial numerator $\mathcal{N}$ is given by
\begin{equation}
\begin{split}
    \mathcal{N} =\;& -\omega_\kappa^7 a^{22} a'^{10} + 42 \theta_R \omega_\kappa^6 a^{19} a'^{10} a'' + 72 \theta_R \omega_\kappa^6 a^{20} a'^8 a''^2 + 864 \theta_R^2 \omega_\kappa^5 a^{17} a'^8 a''^3 \\
    & - 528519168 \theta_R^7 a'^{12} a''^6 + 223948800 \theta_R^7 a a'^{10} a''^7 + 454616064 \theta_R^7 a^2 a'^8 a''^8 \\
    & + 186624 \theta_R^6 a^5 a'^2 a''^7 (524 \omega_\kappa a'^6 + 3 \theta_R a''^4) - 108 \theta_R^3 \omega_\kappa^3 a^{14} a'^2 a''^4 (373 \omega_\kappa a'^6 + 12 \theta_R a''^4) \\
    & - 46656 \theta_R^5 \omega_\kappa a^8 a'^2 a''^6 (-54 \omega_\kappa a'^6 + 13 \theta_R a''^4) - 2592 \theta_R^4 \omega_\kappa^2 a^{11} a'^2 a''^5 (266 \omega_\kappa a'^6 + 27 \theta_R a''^4) \\
    & + 404352 \theta_R^5 \omega_\kappa a^7 a'^4 a''^5 (64 \omega_\kappa a'^6 + 69 \theta_R a''^4) + 2160 \theta_R^3 \omega_\kappa^3 a^{13} a'^4 a''^3 (85 \omega_\kappa a'^6 + 72 \theta_R a''^4) \\
    & + 36 \theta_R^2 \omega_\kappa^4 a^{16} a'^4 a''^2 (155 \omega_\kappa a'^6 + 72 \theta_R a''^4)+ 1296 \theta_R^3 \omega_\kappa^3 a^{12} a'^6 a''^2 (-295 \omega_\kappa a'^6 + 82 \theta_R a''^4) \\
    & - 4478976 \theta_R^6 a^3 a'^6 a''^5 (59 \omega_\kappa a'^6 + 85 \theta_R a''^4)+ 216 \theta_R^2 \omega_\kappa^4 a^{15} a'^6 a'' (-59 \omega_\kappa a'^6 + 146 \theta_R a''^4) \\
    & + 559872 \theta_R^6 a^4 a'^4 a''^6 (212 \omega_\kappa a'^6 + 155 \theta_R a''^4)+ 3 \theta_R \omega_\kappa^5 a^{18} a'^6 (-59 \omega_\kappa a'^6 + 312 \theta_R a''^4) \\
    & - 10368 \theta_R^4 \omega_\kappa^2 a^9 a'^6 a''^3 (590 \omega_\kappa a'^6 + 699 \theta_R a''^4)+ 3888 \theta_R^4 \omega_\kappa^2 a^{10} a'^4 a''^4 (760 \omega_\kappa a'^6 + 827 \theta_R a''^4) \\
    & - 46656 \theta_R^5 a^6 a''^4 (1180 \omega_\kappa^2 a'^{12} + 2120 \theta_R \omega_\kappa a'^6 a''^4 + 33 \theta_R^2 a''^8)\,.
\end{split}
\end{equation}
To consistently quantize these scalar fluctuations, the initial boundary conditions for the Mukhanov-Sasaki modes must be established in the deep sub-horizon regime ($k \gg \mathcal{H}$). In this high-frequency limit, the effective term $z''/z$ becomes negligible relative to the spatial gradient term, and the governing equation of motion reduces asymptotically to that of a harmonic oscillator with a time-varying frequency. To isolate the Bunch-Davies vacuum, we enforce the normalized positive-frequency WKB asymptotic behavior. Consequently, in the asymptotic past ($\eta \to -\infty$), the mode functions satisfy the boundary condition
\begin{equation}
    v_k(\eta) \simeq \frac{1}{\sqrt{2 k c_s(\eta)}} \exp\left( -i k \int^\eta c_s(\tilde{\eta}) \, d\tilde{\eta} \right)\,.
\end{equation}
Setting this initial state enables one to solve the mode evolution through horizon crossing ($k = aH$) where the field amplitudes freeze out, allowing for the calculation of the primordial scalar power spectrum $\mathcal{P}_{\mathcal{R}}(k)$.

\subsubsection{Tensor power spectrum and modified dispersion relations}
\label{subsect:TensorPowerSpectrum}

Let us apply the EFT reduction procedure now to the tensor sector.
%Turning our attention to the tensor sector, we apply an analogous EFT reduction procedure as in the scalar case. 
We consider the pure Starobinsky action as the background theory, and later expand the tensor perturbations, defined by the transverse-traceless spatial metric fluctuations $h_{ij}$, up to second order. To erase higher-order time derivatives into the quadratic action, we perform a perturbative reduction of order substituting the background tensor equations of motion derived from the Starobinsky sector.

Following this reduction, the resulting effective action depends only on the tensor perturbation and its first time derivative. Introducing the Mukhanov-Sasaki variable for the tensor modes, $v_T = z_T h$, the action can be written as
\begin{equation}
    S_{\text{HD}}^{(T)} = \int d\eta \, \frac{1}{2} \left[ v_T'^2 - c_t^2(\eta) k^2 v_T^2 - \alpha(\eta) k^4 v_T^2 + \frac{z_T''}{z_T} v_T^2 \right]\,.
\end{equation}
The emergence of the anomalous $\alpha(\eta) k^4$ term is, of course, a consequence of the higher-order spatial derivatives. The square of the effective speed of sound, $c_t^2(\eta)$, is given by
\begin{equation}
\begin{split}
    c_t^2(\eta) =\;& 1 - \frac{2 \theta_C \left( \omega_\kappa a^3 a'^2 + 12 \theta_R a'^2 a'' + 6 \theta_R a a''^2 \right)^2}{a a'^2 \left( \omega_\kappa a^3 + 12 \theta_R a'' \right)^3} \\
    & + \frac{\omega_C \mathcal{N}_T}{\theta_R a'^4 \left( \omega_\kappa a^4 + 12 \theta_R a a'' \right)^5}\,,
\end{split}
\end{equation}
where the polynomial numerator $\mathcal{N}_T$ is  
\begin{equation}
\begin{split}
    \mathcal{N}_T =\;& -12 \omega_\kappa^5 a^{16} a'^6 - 2006 \theta_R \omega_\kappa^4 a^{13} a'^6 a'' + 192 \theta_R \omega_\kappa^4 a^{14} a'^4 a''^2 \\
    & - 79152 \theta_R^2 \omega_\kappa^3 a^{10} a'^6 a''^2 + 8856 \theta_R^2 \omega_\kappa^3 a^{11} a'^4 a''^3 - 1323648 \theta_R^3 \omega_\kappa^2 a^7 a'^6 a''^3 \\
    & + 153720 \theta_R^3 \omega_\kappa^2 a^8 a'^4 a''^4 + 40663296 \theta_R^5 a'^8 a''^4 + 1190592 \theta_R^4 \omega_\kappa a^5 a'^4 a''^5 \\
    & - 29901312 \theta_R^5 a a'^6 a''^5 + 3473280 \theta_R^5 a^2 a'^4 a''^6 \\
    & + 864 \theta_R^3 \omega_\kappa a^6 a'^2 a''^2 (1961 \omega_\kappa a'^6 + 26 \theta_R a''^4) \\
    & + \theta_R \omega_\kappa^3 a^{12} a'^2 (1961 \omega_\kappa a'^6 + 48 \theta_R a''^4) \\
    & - 3456 \theta_R^4 a^3 a'^2 a''^3 (-3922 \omega_\kappa a'^6 + 105 \theta_R a''^4) \\
    & + 24 \theta_R^2 \omega_\kappa^2 a^9 a'^2 a'' (3922 \omega_\kappa a'^6 + 207 \theta_R a''^4) \\
    & + 432 \theta_R^4 a^4 a''^4 (-23600 \omega_\kappa a'^6 + 291 \theta_R a''^4)\,.
\end{split}
\end{equation}
The higher-order dispersion parameter $\alpha(\eta)$ reads
\begin{equation}
    \alpha(\eta) = \frac{16 \omega_C \left( 16 \omega_\kappa^2 a^6 a'^4 - 3 \omega_\kappa^2 a^7 a'^2 a'' + 384 \theta_R \omega_\kappa a^3 a'^4 a'' - 54 \theta_R \omega_\kappa a^4 a'^2 a''^2 + 2304 \theta_R^2 a'^4 a''^2 - 216 \theta_R^2 a a'^2 a''^3 + 72 \theta_R^2 a^2 a''^4 \right)}{a'^2 \left( \omega_\kappa a^4 + 12 \theta_R a a'' \right)^3}\,.
\end{equation}

Finally, the canonical quantization prescription of these tensor modes requires establishing initial conditions deeply within the sub-horizon regime ($k \gg \mathcal{H}$). Unlike the standard inflationary scenario, the presence of the anomalous $\alpha(\eta) k^4$ term fundamentally modifies the UV dispersion relation. Neglecting the subdominant term $z_T''/z_T$, the effective frequency of the modes in this regime is given by $\omega_k^2(\eta) \approx c_t^2(\eta) k^2 + \alpha(\eta) k^4$. While the $k^4$ contribution dominates in the ultraviolet past ($k \to \infty$), numerical integrations are necessarily starting at a finite conformal time deeply inside the horizon. To ensure a good numerical precision and appropriately select the adiabatic vacuum state, we use the generalized WKB initial condition
\begin{equation}
    v_{T,k}(\eta_i) \simeq \frac{1}{\sqrt{2 \omega_k(\eta_i)}} \exp\left( -i \int^{\eta_i} \omega_k(\eta) \, d\eta \right)\,,
\end{equation}
where the time-dependent frequency is defined as
\begin{equation}
    \omega_k(\eta) = \sqrt{c_t^2(\eta) k^2 + \alpha(\eta) k^4}\,.
\end{equation}
The definition of this vacuum state imposes a bound on the viability of the EFT framework, since we must guaranty that $\omega_k^2 > 0$ throughout the early inflationary phase. In particular, the spatial dispersion parameter $\alpha(\eta)$ must remain strictly positive; otherwise, the frequency would become purely imaginary at high momenta.

To extract the behavior during inflation, we evaluate the effective speeds of sound within the quasi-de Sitter slow-roll regime. Specifically, we focus on the high-energy inflationary plateau where the quadratic Starobinsky term dominates the background dynamics, corresponding to the limit $\omega_\kappa \to 0$. By imposing the standard slow-roll kinematic relations in conformal time, namely $a' = a\mathcal{H}$ and $a'' \simeq a\mathcal{H}^2(2-\epsilon)$, and isolating the leading-order contributions in the limit of vanishing slow-roll parameters ($\epsilon \to 0$), the highly non-linear effective sound speeds simplifies considerably.

In this deep slow-roll plateau limit, the scalar speed of sound squared reduces exactly to unity
\begin{equation}
    c_s^2 \simeq 1\,.
\end{equation}
This indicates that the scalar fluctuations propagate at the speed of light, effectively decoupling from the higher-derivative EFT corrections in this asymptotic regime and recovering the standard inflationary scenario. Conversely, the tensor speed of sound  acquires a modification which is dependent on the higher-order couplings
\begin{equation}
    c_t^2 \simeq 1 - \frac{\theta_C}{3\theta_R} - \frac{37 H^2 \omega_C}{3\theta_R}\,,
\end{equation}
where we have mapped the conformal parameters onto the physical Hubble scale via $H = \mathcal{H}/a$. 

Therefore, the ratio of the tensor-to-scalar sound speeds is determined entirely by the tensor sector structure
\begin{equation}
    \frac{c_t^2}{c_s^2} \simeq 1 - \frac{\theta_C}{3\theta_R} - \frac{37 H^2 \omega_C}{3\theta_R}\,.
\end{equation}

\subsubsection{Numerical implementation and power spectra extraction}
\label{subsect:NumericalImplementation}

\begin{figure}[htbp]
    \centering
    \begin{subfigure}[b]{0.49\textwidth}
        \centering
        \includegraphics[width=\textwidth]{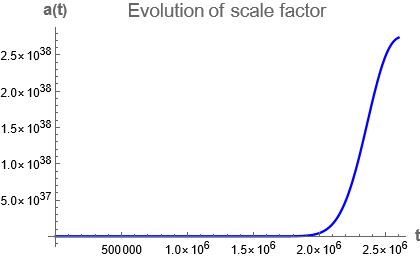}
        \caption{Evolution of the background scale factor $a(t)$.}
        \label{fig:Fig13a}
    \end{subfigure}
    \hfill
    \begin{subfigure}[b]{0.49\textwidth}
        \centering
        \includegraphics[width=\textwidth]{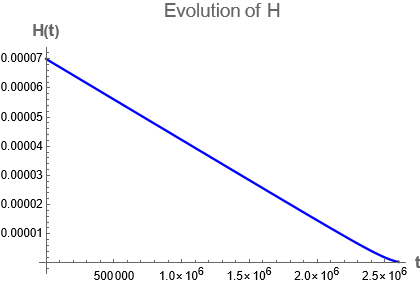}
        \caption{Evolution of the Hubble parameter $H(t)$.}
        \label{fig:Fig13b}
    \end{subfigure}
    \caption{Numerical simulation of the background inflationary dynamics for the baseline Starobinsky model with $\theta_R = 10^9$.}
    \label{fig:Fig13}
\end{figure}

Finally, we numerically integrate the corresponding equations of motion to extract the asymptotic values of the primordial power spectra, which freeze out shortly after the modes exit the Hubble horizon. To achieve this, we must first simulate a background geometry driven by Starobinsky inflation. Observational constraints on the amplitude of primordial scalar perturbations restrict the scalaron mass, $m_s^2 = 1/(6\theta_R)$, to be approximately $m_s \sim 10^{-5} M_{\text{Pl}}$. Consequently, we adopt a parameter value of $\theta_R = 10^9$ for our numerical analysis. The resulting background dynamics is shown in \Cref{fig:Fig13}, where the Hubble parameter exhibits the characteristic slow-roll linear decay, well-approximated analytically by $H(t) \approx H_i - \frac{1}{36\theta_R}(t-t_i)$.

Once the background evolution is fixed, we proceed to numerically solve the mode equations for the respective Mukhanov-Sasaki variables $v_T$ and $v_S$. The dimensionless tensor and scalar power spectra, $\Delta_t$ and $\Delta_s$, are constructed from the frozen-out amplitudes of these Fourier mode functions via the relations
\begin{equation}
    \Delta_t(k) = \frac{k^3}{\pi^2}|h_k(t)|^2 \,, \quad \text{and} \quad \Delta_s(k) = \frac{k^3}{2\pi^2}|\mathcal{R}_k(t)|^2 \,.
\end{equation}
To properly initialize the system in the high-frequency sub-horizon regime ($k \gg \mathcal{H}$), we use the generalized WKB initial state derived in the previous section. Accounting for the modified ultraviolet dispersion relation introduced by the higher-order corrections, the mode integration is initialized deep inside the horizon via
\begin{equation}
    v_{T,k}(\eta_i) \simeq \frac{1}{\sqrt{2 \omega_k(\eta_i)}}\,.
\end{equation}

\begin{figure}[H]
    \centering
    \begin{subfigure}[b]{0.49\textwidth}
        \centering
        \includegraphics[width=\textwidth]{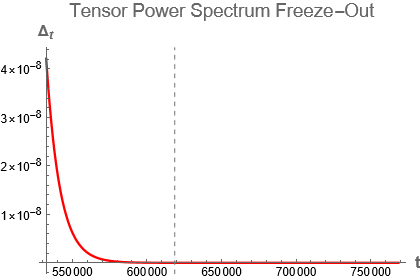}
        \caption{Power spectrum for the tensor perturbations $\Delta_t(k)$.}
        \label{fig:Fig14a}
    \end{subfigure}
    \hfill
    \begin{subfigure}[b]{0.49\textwidth}
        \centering
        \includegraphics[width=\textwidth]{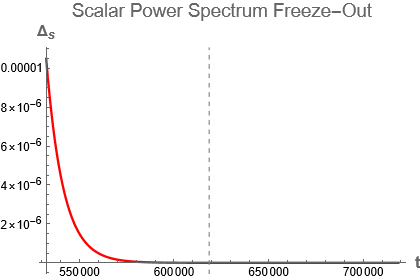}
        \caption{Power spectrum for the scalar perturbations $\Delta_s(k)$.}
        \label{fig:Fig14b}
    \end{subfigure}
    \caption{Time evolution and subsequent super-horizon freeze-out of the primordial power spectra for a comoving scale $k = 1.59 \times 10^{12}$. The higher-derivative coupling parameters are set to $\theta_C = -\frac{\theta_R}{20}$ and $\omega_C = \frac{\theta_R}{100}$. The vertical dashed grey line marks the exact moment of horizon crossing ($k = aH$).}
    \label{fig:Fig14}
\end{figure}

A representative example of this evolution is presented in \Cref{fig:Fig14}, which displays the numerical solutions for the tensor and scalar power spectra in plots \labelcref{fig:Fig14a} and \labelcref{fig:Fig14b}, respectively. This simulation corresponds to a specific comoving scale $k$ that exits the horizon $N = 50$ $e$-folds before the end of inflation. The plot explicitly shows the sub-horizon dynamic behavior of the modes, followed by the characteristic freeze-out of the power spectra immediately after horizon crossing, which successfully yields a constant super-horizon amplitude.

\begin{figure}[htbp]
    \centering
    \begin{subfigure}[c]{0.42\textwidth}
        \centering
        \includegraphics[width=\textwidth]{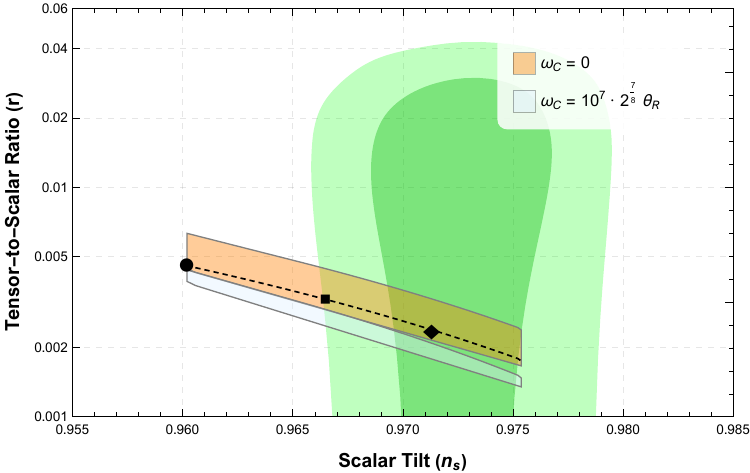} % Replace with your first plot filename
        \label{fig:omega_c_variation}
    \end{subfigure}
    \hfill
    \begin{subfigure}[c]{0.42\textwidth}
        \centering
        \includegraphics[width=\textwidth]{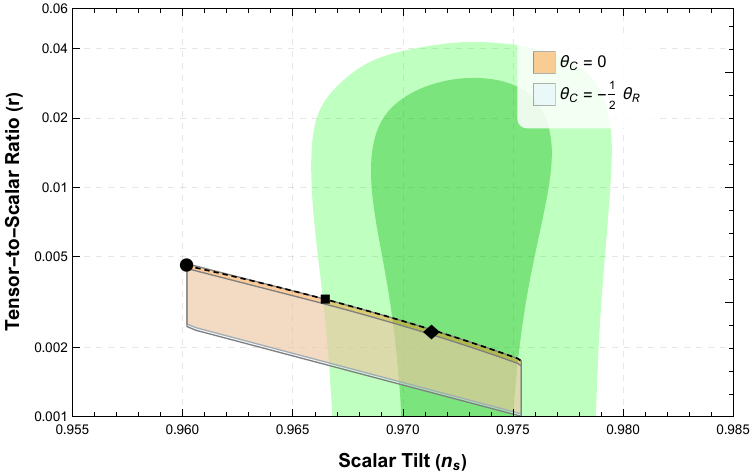} % Replace with your second plot filename
        \label{fig:theta_c_variation}
    \end{subfigure}
     \hfill
    \begin{subfigure}[t]{0.13\textwidth}
        \vspace{-2.3cm}
        \includegraphics[width=\textwidth]{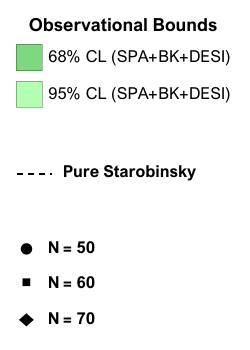}
    \end{subfigure}
    \caption{Contrast of theoretical prediction with observational bounds  in the $(n_s, r)$ plane. Plot (a) on the left illustrates the accessible parameter space domain when varying $\theta_C$ and the number of e-folds $N$ for fixed values of $\omega_C$, whereas plot (b) on the right shows the converse, varying $\omega_C$ at fixed values of $\theta_C$. In both plots, the orange domain  represents the baseline model space where the respective fixed parameter is set to zero. The light blue domain highlights the shifting of the predicted parameter space under non-zero EFT parameter configurations. The dashed black line marks the standard pure Starobinsky inflation prediction, with solid markers indicating $N = 50$, $60$, and $70$ $e$-folds. The background green shaded areas denote the 68\% and 95\% CL observational bounds from SPA+BK+DESI \cite{Balkenhol:2025wms}.}
    \label{fig:eft_constraints}
\end{figure}

Finally, with the numerically evaluated power spectra at horizon exit, we extract the primary cosmological observables, namely the scalar spectral index $n_s$ and the tensor-to-scalar ratio $r$. In order to map the phenomenological implications of the EFT, we compute these quantities across a wide domain of the higher-derivative coupling parameter space. The theoretical predictions arising from this exploration are presented in \Cref{fig:eft_constraints}. This figure illustrates how the inclusion of sixth-order operators induces different shifts in the $n_s$-$r$ plane relative to the standard Starobinsky model, facilitating a direct comparison with current observational constraints from the CMB.

In order to probe the upper limits of the phenomenological deviations allowed within this framework, the parameter ranges for the higher-derivative EFT operators are chosen such that their contributions become comparable in magnitude to the fundamental Starobinsky term. Given the quasi-de Sitter inflationary background, the relevant terms in the action scale structurally as
\begin{equation}
    \theta_R H^4 + \theta_C H^4 + \omega_C H^6\,,
\end{equation}
where the physical Hubble parameter $H$ is of order $10^{-5} M_{\text{Pl}}$, as established by the scalar amplitude normalization. For the Weyl-squared $C^2$ operator to yield an energy contribution of the same order as the quadratic Ricci sector, the respective coupling parameter must scale as $\theta_C \sim \theta_R$. Conversely, matching the contribution of the six-derivative $C\Box C$ operator requires that
\begin{equation}
    \omega_C \sim \frac{\theta_R}{H^2} \approx 2 \times 10^{17}\,.
\end{equation}
Consequently, while the numerical value of the $\omega_C$ coupling parameter is hierarchically larger than $\theta_R$, its physical contribution is dynamically modulated by the inflationary Hubble scale, ensuring that both higher-derivative sectors operate at the same effective energy threshold during the horizon-exit epoch.
% Placeholder for the figure environment

As shown in \Cref{fig:eft_constraints}, the primary effect of the EFT corrections is a vertical shift in the predicted parameter space, significantly modifying the tensor-to-scalar ratio, $r$, while leaving the scalar tilt, $n_s$, largely unaffected. Specifically, when $\omega_C$ is fixed, introducing a positive $\theta_C$ increases $r$, whereas a negative $\theta_C$ decreases it. Conversely, varying $\omega_C$ at a fixed $\theta_C$ systematically drives the value of $r$ downwards. Meanwhile, the horizontal placement along these modified trajectories is almost entirely dictated by the number of $e$-folds, $N$. Just as in the standard Starobinsky model, increasing $N$ from 50 to 70 shifts the scalar tilt toward higher values, driving the predictions horizontally toward the center of the contour bounds. From a phenomenological perspective, the combination of these EFT parameters and a sufficiently high number of $e$-folds can push the theoretical predictions perfectly into the highly favored 68\% CL domain of the SPA+BK+DESI data.

\newpage
\section{Conclusions}
In this work we have presented a comprehensive cosmological analysis of sixth-order higher-derivative gravity, extending several well-known results of quadratic gravity to the super-renormalizable regime. The theory exhibits a significantly richer cosmological structure than Einstein gravity and its quadratic extension, both at the level of homogeneous solutions and in the dynamics of cosmological perturbations.

For homogeneous and isotropic backgrounds, we find %the generalized Friedmann equations and analyzed the resulting solution space.
besides the standard expanding solutions, also familiar in General Relativity,  recollapsing cosmologies and non-singular bouncing solutions that are considerably more restricted in quadratic gravity. By exploiting the Einstein-frame representation, we showed explicitly how the additional higher-derivative degrees of freedom modify the main conditions of the Hawking-Penrose singularity theorems, providing the mechanism responsible for the existence of regular cosmological evolutions. We also point out the connection of the mechanism that avoids the singularities and the excitation of ghost degrees of freedom inherent to local higher-derivative theories.

%At the same time, the analysis makes explicit the intimate relation between singularity avoidance and 
We derive the complete equations governing tensor perturbations around Minkowski, radiation-dominated, matter-dominated and de Sitter backgrounds. Their analytical asymptotic solutions, together with numerical integrations, allowed us to identify the domains of parameter space where the perturbations remain bounded. The resulting stability conditions are directly related to the structure of the propagator. While the infrared limit smoothly reproduces the predictions of General Relativity, the ultraviolet regime is dominated by the additional massive modes associated with the sixth-order operators. An important result of this analysis is that all singularity-free cosmological solutions found in the present theory develop unstable tensor perturbations, suggesting that their regular background evolution is not dynamically robust within the local sixth-order framework.

We also analyzed the scalar perturbation sector and subsequently implemented the effective field theory reduction-of-order formalism. This approach removes the spurious higher-order solutions while consistently retaining the leading physical corrections induced by the higher-derivative operators. Within this perturbative framework, we computed the corrections to the primordial scalar and tensor power spectra generated during inflation and evaluated the corresponding predictions for the scalar spectral index and the tensor-to-scalar ratio. The sixth-order operators lead to well-controlled departures from the Starobinsky model, shifting the theoretical predictions within the $(n_s,r)$ plane while preserving the validity of the effective field theory expansion.

In summary, the results show that sixth-order higher-derivative gravity provides a natural extension of quadratic gravity with a substantially richer phenomenology. Although the complete theory continues to exhibit the usual difficulties associated with ghost degrees of freedom and the classical instability of gravitational backgrounds, the effective field theory treatment remains predictive and allows a connection of ultraviolet gravitational corrections   with observable inflationary signatures. %Future work could extend 
The present analysis can be extended to vector perturbations, nonlinear stability or reheating dynamics which can be the subject for future work. Also, it would be worthwhile to determine whether the alternative prescriptions proposed in the literature to handle Ostrogradsky ghosts, such as Lee-Wick-type contour deformations, fakeons, non-locality or ghost-condensate approaches, can stabilize the singularity-free backgrounds while preserving their singularity-avoiding properties.

\end{document}